\documentclass[reprint,amsmath,amssymb,aps,pra, superscriptaddress]{revtex4-1}
\usepackage{graphicx}
\usepackage{amsmath,amssymb,cleveref}
\usepackage{url}
\usepackage{layout}
\usepackage{color}
\usepackage{braket}
\usepackage{physics}
\usepackage{mathrsfs}
\usepackage{booktabs}
\usepackage{listings}
\usepackage{dcolumn}
\usepackage{bm}
\usepackage{overpic}

\def\lesssim{\ \raise.3ex\hbox{$<$}\kern-0.8em\lower.7ex\hbox{$\sim$}\ }
\def\gesim{\ \raise.3ex\hbox{$>$}\kern-0.8em\lower.7ex\hbox{$\sim$}\ }

\begin{document}
\title{Dimensionality of a strongly interacting 2D-3D Fermi-Fermi mixture from the perspective of superfluid instability and excitation properties}
\author{Haruka Takeda}
\author{Saki Hirai}
\affiliation{Department of Physics, Keio University, 3-14-1 Hiyoshi Kohoku-ku, Yokohama 223-8522, Japan}
\author{Shumpei Iwasaki}
\affiliation{Institute for Solid State Physics, University of Tokyo, Kashiwa, Chiba 277-8581, Japan}
\author{Yoji Ohashi}
\affiliation{Department of Physics, Keio University, 3-14-1 Hiyoshi Kohoku-ku, Yokohama 223-8522, Japan}

\begin{abstract}
    We theoretically investigate strong-coupling properties of an attractively interacting Fermi atomic gas, 
    where the Cooper-pair formation occurs between atoms belonging to different dimensional bands.
    Including pairing fluctuations within the framework of the self-consistent $T$-matrix approximation (SCTMA), 
    we examine how the BCS-type superfluid phase transition temperature $T_\mathrm{c}$ varies as one moves from the 3D-3D to the 2D-3D system, 
    in the wide parameter region with respect to the strength of the pairing interaction. 
    In the 2D-3D limit, we find that, 
    while the mean-field BCS theory predicts $T_\mathrm{c}>0$ in the strong-coupling regime, 
    $T_\mathrm{c}$ is remarkably suppressed down to zero by pairing fluctuations that are strongly enhanced by the mixed-dimensionality of the system. 
    As the origin of this, 
    we clarify that the lower-dimensional (2D) component dominates the superfluid instability, 
    so that the vanishing $T_\mathrm{c}$ is the same phenomenon as that in the 2D-2D case.
    We also point out that this can already be seen in the mean-field level, 
    when one examines the propagation of the Goldstone mode. 
    On the other hand, 
    we find that the pseudogap phenomenon, which is known as a precursor of Cooper-pair formation, 
    exhibits a 3D character of the 2D-3D system.
    These results indicate that the dimensionality of a strongly interacting Fermi gas depends on what we observe.
\end{abstract}
\maketitle

 \par
 \section{Introduction}
 Ultracold atomic gases have significantly contributed to our understanding of various quantum many-body systems~\cite{RevModPhys.80.885,natphys.8.267,RevModPhys.80.1215,Randeria2014,OHASHI2020103739,natrevphys.2.411}.
 In particular, the observation of the BCS (Bardeen-Cooper-Schrieffer)-BEC (Bose-Einstein condensate) crossover is one of the most exciting achievements in cold Fermi gas physics~\cite{PhysRevLett.92.040403,PhysRevLett.92.120403,PhysRevLett.92.120401,PhysRevLett.93.050401,nature.435.1047}, 
 and  has boosted experimental/theoretical interests in strongly interacting Fermi-Fermi mixtures~\cite{PhysRevA.71.012708,PhysRevA.74.063628,PhysRevLett.102.230402,science.354.96,Tempere2009,Klimin2012}.
 While atomic gases were first examined in the three-dimensional continuum case, 
 we can now also deal with various band systems, 
 as well as lower-dimensional cases, 
 thanks to great experimental efforts on the optical lattice technique~\cite{PhysRevLett.87.160405,PhysRevLett.91.250402,PhysRevLett.92.190401,science.305.1125,nature.429.227,PhysRevLett.103.140401,PhysRevLett.114.230401,Masaki2008}.

 This technique has recently further developed to the so-called species-selective optical lattice~\cite{PhysRevA.74.013616,PhysRevA.75.053612,PhysRevLett.104.153202,PhysRevA.98.051602}. 
 Using this, experimental groups have realized mixed-dimensional atomic gases~\cite{PhysRevLett.104.153202,PhysRevA.98.051602}, 
 where different kinds of atoms move in different dimensions.
 At present, such a dimensional imbalance has been realized in $^{41}$K-$^{87}$Rb~\cite{PhysRevLett.104.153202} and $^{7}$Li-$^{174}$Yb~\cite{PhysRevA.98.051602} Bose-Bose gases, as well as a $^{7}$Li-$^{173}$Yb~\cite{PhysRevA.98.051602} Bose-Fermi gas, 
 although a mixed-dimensional Fermi-Fermi gas has not been realized yet. 
 However, stimulated by these recent experimental progress, 
 physical properties of the Fermi-Fermi case have also attracted much attention theoretically~\cite{PhysRevLett.101.170401,PhysRevA.82.063628,epjb.83.445,PhysRevA.95.053633}.

\begin{figure}
  \centering
  \includegraphics[width=0.9\linewidth]{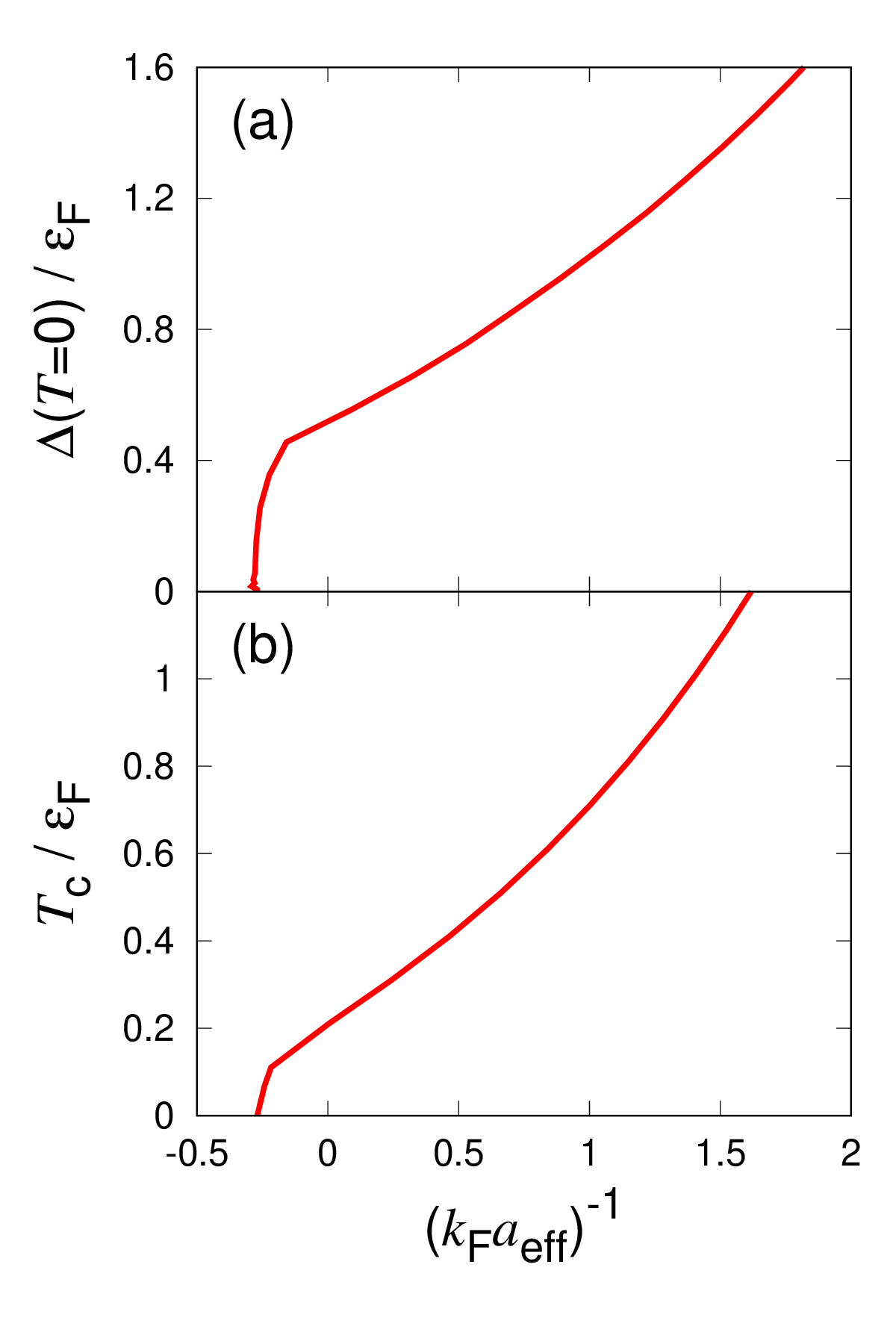}
  \vspace*{-0.5cm}
  \caption{
    Mean-field BCS results for 
    (a) superfluid order parameter $\Delta$ at $T=0$, and 
    (b) superfluid phase transition temperature $T_\mathrm{c}$, 
    in a two-component 2D-3D Fermi-Fermi mixture with an $s$-wave pairing interaction. 
    $\epsilon_\mathrm{F}=k_\mathrm{F}^2/(2m)$ and $k_\mathrm{F}$ is given in Eq.~\eqref{eq:k_F}, 
    where $m$ is an atomic mass. 
    $a_\mathrm{efff}$ is the effective $s$-wave scattering length in Eq.~\eqref{eq:a_eff}.
    We have used the mean-field theory in Ref.~\cite{PhysRevA.82.063628} to obtain these results, 
    and the outline of this theory is explained in Appendix \ref{sec:mf_BCS}.}
  \label{fig:gap_v_ng}
 \end{figure}

 Recently, Iskin and Suba\c{s}\i\ studied an attractively interacting mixed-dimensional two-component Fermi gas within the mean-field BCS approximation, 
 to draw the ground state phase diagram in terms of the interaction strength and the dimensional imbalance~\cite{PhysRevA.82.063628}.
 Particularly in the case of a 2D-3D mixture, they clarified that, 
 while the BCS-type superfluid phase transition vanishes in the weak-coupling regime ($(k_\mathrm{F}a_\mathrm{eff})^{-1}\lesssim -0.3$, 
 where $k_\mathrm{F}$ is the Fermi wavelength and $a_\mathrm{eff}$ is the effective $s$-wave scattering length given in Eqs.~\eqref{eq:k_F} and \eqref{eq:a_eff}) 
 because of the mismatch of the 2D and 3D Fermi surfaces, 
 a strong pairing interaction overcomes this depairing effect, 
 giving the superfluid phase transition when $(k_\mathrm{F}a_\mathrm{eff})^{-1} \gtrsim -0.3$.
 Indeed, calculating the superfluid order parameter $\Delta$ at $T=0$ within the mean-field theory in Ref.~\cite{PhysRevA.82.063628}, we obtain $\Delta>0$ in the strong-coupling side, as shown in Fig.~\ref{fig:gap_v_ng}(a).

 Although the mean-field study in Ref.~\cite{PhysRevA.82.063628} clarified how the Fermi surface mismatch affects the superfluid phase transition in a mixed-dimensional Fermi gas, 
 the importance of mixed-dimensional pairing fluctuations still remains to be elucidated.
 Regarding this, we recall the Hohenberg-Mermin-Wagner theorem~\cite{PhysRevLett.17.1133,PhysRev.158.383}, 
 stating that pairing fluctuations completely destroy superfluid long-range order at $T>0$ in the two-dimensional system.
 Thus, it is an interesting problem to examine in the 2D-3D case whether, even when pairing fluctuations are taken into account, 
 the superfluid transition temperature $T_\mathrm{c}$ obtained in the mean-field theory (see Fig.~\ref{fig:gap_v_ng} (b)) still remains non-zero due to the ``3D'' character of the system, or 
 the BCS-type long-range order is completely suppressed when $T>0$ due to the ``2D'' component.

 In this paper, we theoretically investigate strong-coupling effects on the BCS-type superfluid phase transition in a mixed-dimensional Fermi atomic gas, 
 where Cooper pairs are formed between atoms having different dimensional bands.
 For this purpose, we include pairing fluctuations within the framework of self-consistent $T$-matrix approximation (SCTMA)~\cite{PhysRevB.49.12975,haussmann2003self,PhysRevA.75.023610,PhysRevLett.100.140407,PhysRevA.82.013624,ENSS2011770,PhysRevA.90.043622,Kagamihara2019}, 
 to evaluate $T_\mathrm{c}$ as a function of the interaction strength and the dimensional imbalance.
 In addition, we also examine effects of mixed-dimensional pairing fluctuations from the viewpoint of the pseudogap phenomenon appearing in single particle excitations.

 Here, we comment on the reason why we employ SCTMA in this paper: In cold Fermi gas physics, besides SCTMA, various strong-coupling theories, such as the Nozi\`eres-Schmitt-Rink (NSR) theory~\cite{Nozieres1985, PhysRevA.67.063612,PhysRevLett.71.3202,liu_bcs-bec_2006,PhysRevA.109.063309}, 
 $T$-matrix approximation (TMA)~\cite{PhysRevB.70.094508,physrep.412.1,PhysRevLett.102.190402,PhysRevLett.104.240407,PhysRevA.84.043647,PhysRevB.85.024517,PhysRevA.83.053623,PhysRevA.88.013637,PhysRevLett.108.080401,PhysRevA.86.063601}, 
 as well as extended $T$-matrix approximation (ETMA)~\cite{PhysRevA.88.053621,PhysRevA.89.033617,piniFermiGasThroughout2019,PhysRevA.86.043622,PhysRevA.88.053621,PhysRevA.89.013618,PhysRevA.82.043630}, 
 have also extensively been used to study physical properties in the BCS-BEC crossover regime~\cite{RevModPhys.80.1215,Randeria2014,OHASHI2020103739}.
 Although all these theories can describe the whole BCS-BEC crossover region at least qualitatively in the ``balanced'' system, 
 it is known that, except for SCTMA, 
 they sometimes give unphysical results in the presence of spin- and mass-``imbalance''~\cite{PhysRevA.88.053621,PhysRevLett.100.140407,PhysRevA.82.013624,ENSS2011770,PhysRevA.89.033617,piniFermiGasThroughout2019,PhysRevA.86.043622,PhysRevA.90.043622,PhysRevA.88.053621}.
 Indeed, we have numerically confirmed that we also meet the same problem in the dimensional-imbalance case when one employs NSR, TMA and ETMA (although we don't explicitly show the result here).
 On the other hand, we will later show that SCTMA does not encounter such difficulties as one moves from the 3D-3D case to the 2D-3D limit, 
 at least for physical quantities which we are considering in this paper.
 Thus, throughout this paper, we examine the dimensional-imbalanced system by using this strong-coupling scheme.
 
 This paper is organized as follows:
 In Sec.~\ref{sec:Formulation}, we explain our SCTMA formalism for a two-component Fermi gas with dimensional imbalance.
 We show in Sec.~\ref{sec:Tc}, how $T_\mathrm{c}$ varies as one moves from the 3D-3D case to the 2D-3D case, from the weak- to strong-coupling regime.
 In Sec.~\ref{sec:DOS}, we examine strong-coupling corrections to single-particle excitations.
 Throughout this paper, we set $\hbar=k_\mathrm{B}=1$, and the system volume $V$ is taken to be unity.
 
 \section{Formulation}\label{sec:Formulation}
  \subsection{Model mixed-dimensional Fermi gas}
    
  We consider a two-component Fermi gas with a tunable $s$-wave pairing interaction, being loaded on a species-selective one-dimensional optical lattice in the $z$-direction.
  The model Hamiltonian is given~\cite{PhysRevA.82.063628}
  \begin{align}
  H=&\sum_{\bm{p},\sigma=\uparrow,\downarrow}\xi_{\bm{p},\sigma}c^\dagger_{\bm{p},\sigma}c_{\bm{p},\sigma} \notag \\
    &-U\sum_{\bm{p},\bm{p'},\bm{q}}c^\dagger_{\bm{p+q}/2,\uparrow}c^\dagger_{\bm{-p+q}/2,\downarrow}c_{\bm{-p'+q}/2,\downarrow}c_{\bm{p'+q}/2,\uparrow}, \label{eq:Hamiltonian}
  \end{align}
  where $c^\dagger_{\bm{p},\sigma}$ is the creation operator of a Fermi atom with momentum $\bm{p}$, pseudospin $\sigma = \uparrow,\downarrow$, and the kinetic energy,
  \begin{align}
      \xi_{\bm{p},\sigma}&=\epsilon_{\bm{p},\sigma}-\mu_\sigma, \\
      \epsilon_{\bm{p},\sigma}&=\frac{p_x^2+p_y^2}{2m}+2t_\sigma[1-\cos(p_zd_z)], \label{eps}
  \end{align}
  measured from the chemical potential $\mu_\sigma$.
  Here, $t_\sigma$ describes the nearest-neighbor hopping between two-dimensional sheets stacking in the $z$-direction with the spacing $d_z$.

\begin{figure}
   \centering
   \includegraphics[width=0.7\linewidth]{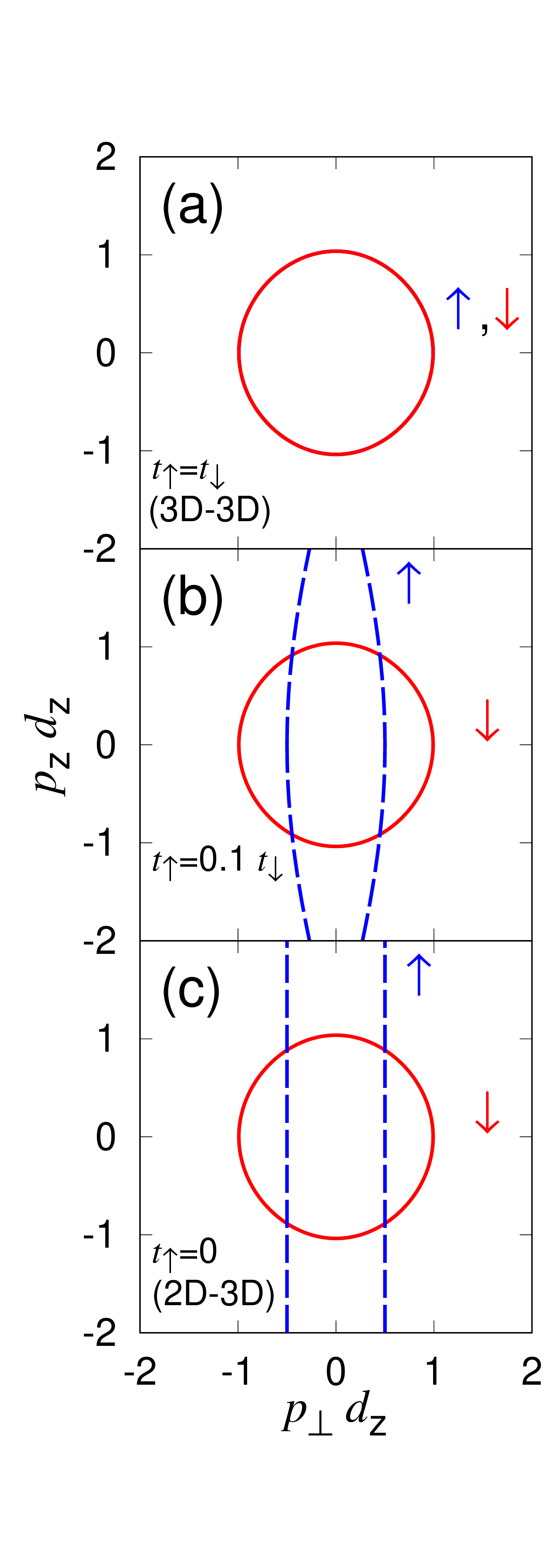}
   \vspace*{-1cm}
   \caption{
     Difference of Fermi surfaces between spin-$\uparrow$ (dashed line) and spin-$\downarrow$ (solid line) components in a non-interacting mixed-dimensional Fermi gas with the kinetic energy in Eq.~\eqref{eps}.
     $p_\perp$ and $p_z$ denote the momentum in the in-plane direction and the inter-layer direction, respectively.
     We set $N_\uparrow=N_\downarrow=1/(6\pi^2d_z^3)$.
   }
   \label{fig:FS}
 \end{figure}

  In Eq.~\eqref{eps}, effects of mixed-dimension are conveniently tuned by the species-dependent hopping parameter $t_\sigma$: 
  In this paper, we fix $t_\downarrow$ and adjust $t_\uparrow$ from $t_\uparrow=t_\downarrow$ to $t_\uparrow=0$.
  Then the system continuously changes from the 3D-3D case ($t_\uparrow=t_\downarrow$) to the 2D-3D limit ($t_\uparrow=0, t_\downarrow\neq 0$), as shown in Fig.~\ref{fig:FS}.
  In Eq.~\eqref{eq:Hamiltonian}, $-U\ (<0)$ is a contact-type $s$-wave pairing interaction.
  To remove the ultraviolet divergence coming from this interaction, we measure the interaction strength in terms of the inverse of the ``effective'' $s$-wave scattering length $a_{\mathrm{eff}}$ given by~\cite{PhysRevA.82.063628}
  \begin{align}
      \frac{4\pi a_\mathrm{eff}}{m}
      &=\frac{-U}{1-U\sum_{\bm{p}}\frac{1}{\epsilon_{\bm{p},\uparrow}+\epsilon_{\bm{p},\downarrow}}} \notag \\
      &=\frac{-U}{1-U\int_{-\pi/d_z}^{\pi/d_z}\frac{\dd{p_z}}{2\pi}\int_0^\infty\frac{\dd{p_\perp}}{(2\pi)^2}2\pi p_\perp\frac{1}{\epsilon_{\bm{p},\uparrow}+\epsilon_{\bm{p},\downarrow}}}.
      \label{eq:a_eff}
  \end{align}
  Eq.~\eqref{eq:a_eff} is a natural extension of the ordinary $s$-wave scattering length $a_s$~\cite{Randeria2014,OHASHI2020103739,RevModPhys.80.1215} in the 3D continuum case to the present dimensional imbalanced one.
  In this scale, the weak-coupling BCS regime and the strong-coupling BEC regime are, respectively, characterized by $(k_\mathrm{F}a_\mathrm{eff})^{-1}\lesssim-1$ and $1\lesssim(k_\mathrm{F}a_\mathrm{eff})^{-1}$, 
  where 
  \begin{align}
    k_\mathrm{F}=(3\pi^2N)^{\frac{1}{3}} \label{eq:k_F}
  \end{align}
  is the Fermi momentum of an assumed free Fermi gas in the absence of optical lattice, 
  with $N$ being the total number of Fermi atoms.
  The region between the two is sometimes referred to as the BCS-BEC crossover region in the literature\cite{RevModPhys.80.1215,Randeria2014,OHASHI2020103739}.
  
  \subsection{Self-consistent $T$-matrix approximation}
\begin{figure}
      \centering
      \includegraphics[width=\linewidth]{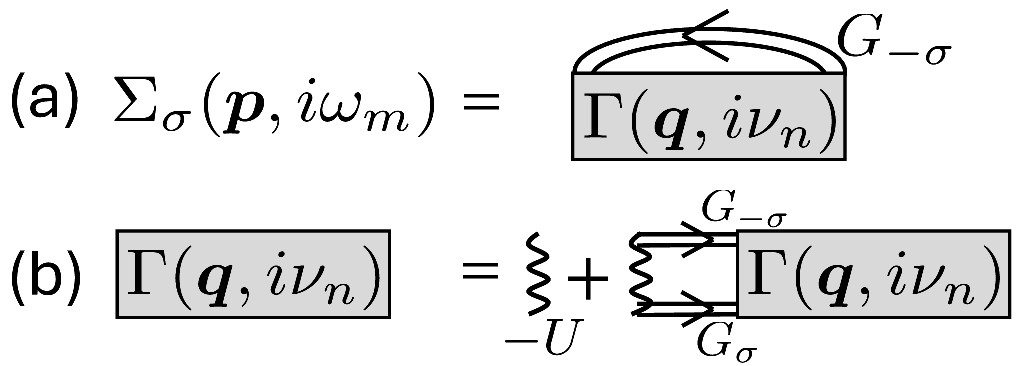}
      \caption{
      (a) SCTMA self-energy $\Sigma_\sigma(\bm{p},i\omega_m)$ in Eq.~\eqref{eq:Self_energy}.
      The double solid line is the dressed Green's function $G_\sigma$ in Eq.~\eqref{eq:Greens_func}.
      (b) Scattering matrix $\Gamma(\bm{q},i\nu_n)$ in Eq.~\eqref{eq:Scattering_matrix}.
      The wavy line describes the paring interaction $-U$.
      In this figure, pseudospin $-\sigma$ represents the opposite spin component to $\sigma=\ \uparrow, \downarrow$.
      }
      \label{pic:Feynman_diagram}
  \end{figure}

  Strong-coupling corrections to single-particle excitations are conveniently incorporated into the theory through the self-energy $\Sigma_\sigma(\bm{p},i\omega_m)$ correction in the single-particle thermal Green's function, given by 
  \begin{align}
      G_\sigma(\bm{p},i\omega_m)=\frac{1}{i\omega_m-\xi_{\bm{p},\sigma}-\Sigma_\sigma(\bm{p},i\omega_m)},\label{eq:Greens_func}
  \end{align}
  where $\omega_m$ is the fermion Matsubara frequency.
  In SCTMA~\cite{PhysRevA.90.043622}, $\Sigma_\sigma(\bm{p},i\omega_m)$ is diagrammatically described as Fig.~\ref{pic:Feynman_diagram} (a), which gives
  \begin{align}
  \Sigma_\sigma(\bm{p},i\omega_m)=T\sum_{\bm{q},i\nu_n}\Gamma(\bm{q},i\nu_n)G_{-\sigma}(\bm{q-p},i\nu_n-i\omega_m). \label{eq:Self_energy}
  \end{align}
  Here, $\nu_n$ is the boson Matsubara frequency, and pseudospin $-\sigma$ denotes the opposite component to $\sigma=\ \uparrow,\downarrow$.
  $\Gamma(\bm{q},i\nu_n)$ is the particle-particle scattering matrix describing effects of fluctuations in the Cooper channel.
  In SCTMA, it is diagrammatically given by Fig.~\ref{pic:Feynman_diagram} (b), from which we have
  \begin{align}
      \Gamma(\bm{q},i\nu_n)=\frac{-U}{1-U\Pi(\bm{q},i\nu_n)}, \label{eq:Scattering_matrix}
  \end{align}
  where
  \begin{align}
      \Pi(\bm{q},i\nu_n)=T\sum_{\bm{p},i\omega_m}&G_\uparrow(\bm{p+q/2},i\nu_n+i\omega_m)\notag \\
      &\times G_\downarrow(\bm{-p+q/2},-i\omega_m),
      \label{eq:PI}
  \end{align}
  is the pair correlation function.

  The BCS-type superfluid transition temperature $T_\mathrm{c}$ is determined from the Thouless criterion~\cite{THOULESS1960553}, 
  stating that the superfluid instability occurs when the particle-particle scattering matrix $\Gamma(\bm{q},i\nu_n)$ has a pole at $\bm{q}=\nu_n=0$.
  Using this, we obtain the $T_\mathrm{c}$-equation as 
  \begin{align}
      1-U\Pi(\bm{q=0},i\nu_n=0)=0. \label{eq:thouless_criterion}
  \end{align}
  We solve Eq.~\eqref{eq:thouless_criterion}, 
  together with the equations for the number $N_\sigma$ of Fermi atoms in the $\sigma$ component, 
  \begin{align}
      N_\sigma=T\sum_{\bm{p},i\omega_m}G_\sigma(\bm{p},i\omega_m)e^{i\omega_m\delta}, \label{eq:Number}
  \end{align}
  where $\delta$ is an infinitesimally small positive number.

  For later convenience, we also introduce the single-particle density of states (DOS) in the $\sigma$-component, 
  \begin{align}
      \rho_\sigma(\omega)
      &=\sum_{\bm{p}}A_\sigma(\bm{p},\omega), \label{eq:rho_omega}
  \end{align}
  as well as the $p_z$-resolved DOS, 
  \begin{align}
      \rho_\sigma(p_z,\omega)
      &=\sum_{\bm{p}_\perp=(p_x,p_y)}A_\sigma(\bm{p},\omega), \label{eq:rho_pz}
  \end{align}
  where 
  \begin{align}
      A_\sigma(\bm{p},\omega)
      &=-\frac{1}{\pi}\mathrm{Im}[G_\sigma(\bm{p},i\omega_m\rightarrow\omega+i\delta)] \label{eq:Ap}
  \end{align}
  is the single-particle spectral weight, 
  which is obtained from the analytic continued single-particle Green's function $G_\sigma(\bm{p},i\omega_m\rightarrow \omega+i\delta)$. 
  We will use these quantities when we discuss the pseudogap phenomenon caused by mixed-dimensional pairing fluctuations in Sec.~\ref{sec:DOS}.

  Before ending this section, we summarize computational information: 
  We numerically solve the coupled $T_\mathrm{c}$-equation \eqref{eq:thouless_criterion} with the two number equations ($\sigma=\uparrow, \downarrow$) in Eq.~\eqref{eq:Number}, 
  to self-consistently determine $(T_\mathrm{c},\mu_\uparrow, \mu_\downarrow)$ for a given parameter set $((k_\mathrm{F}a_\mathrm{eff})^{-1},N_\uparrow,N_\downarrow)$.
  In this paper, we only deal with the spin-balanced case, by setting $N_\uparrow=N_\downarrow=N/2$, 
  where $N$ is the total number of Fermi atoms.
  For $N\equiv k_\mathrm{F}^3/(3\pi^2)$, 
  we set $d_z=k_\mathrm{F}^{-1}$.
  We also set $t_\downarrow/\epsilon_\mathrm{F}=1$, 
  where $\epsilon_\mathrm{F}=k_\mathrm{F}^2/(2m)$.
  In this case, 
  one obtains a nearly spherical Fermi surface in the 3D-3D case ($t_\uparrow=t_\downarrow$) of a free Fermi gas, 
  as shown in Fig.~\ref{fig:FS} (a).
  For the analytic continuation in Eq.~\eqref{eq:Ap}, 
  we employ the Pad\'e approximation~\cite{Vidberg1977}.
  
 \section{Superfluid phase Transition and Effects of mixed-dimension}\label{sec:Tc}
 
\begin{figure}[t]
    \centering
    \includegraphics[width=\linewidth]{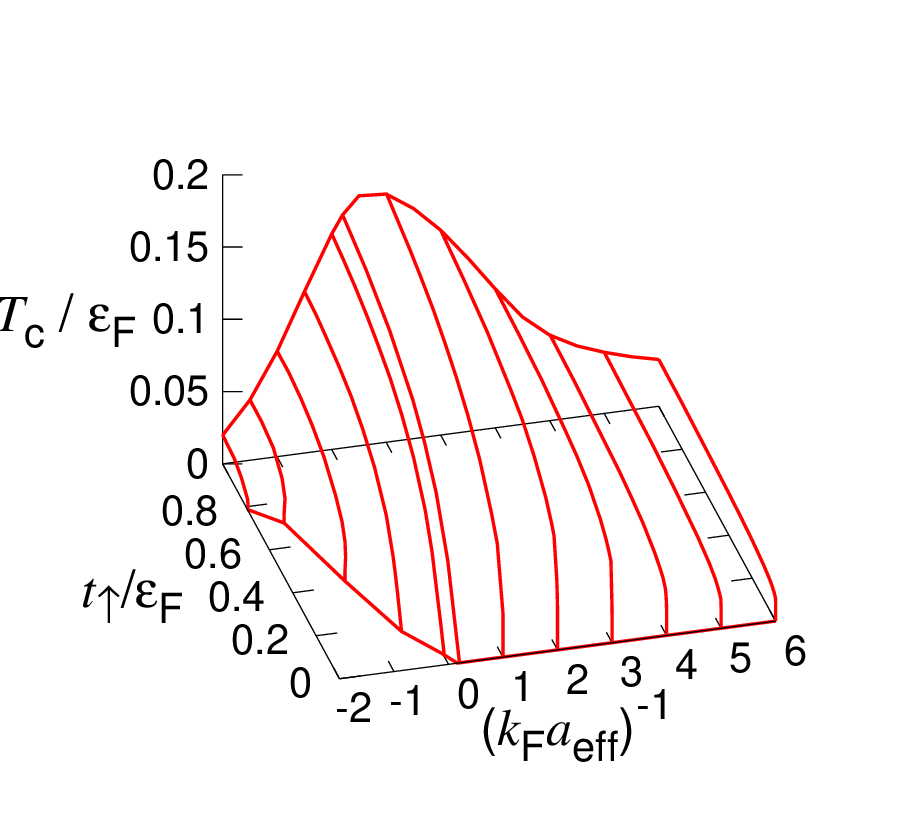}
    \caption{
    Calculated BCS-type superfluid phase transition temperature $T_\mathrm{c}$ in SCTMA, 
    as a function of the interaction strength $(k_\mathrm{F}a_\mathrm{eff})^{-1}$ and the hopping parameter $t_\uparrow$.
    In this figure, 3D-3D and 2D-3D cases are, respectively, realized when $t_\uparrow/\epsilon_\mathrm{F}=1$ and $0$.
    }
    \label{fig:Tc_splot}
 \end{figure}

 Fig.~\ref{fig:Tc_splot} shows the SCTMA result for $T_\mathrm{c}$ in a mixed-dimensional Fermi gas.
 In this figure, the vanishing $T_\mathrm{c}$ seen in the weak-coupling BCS side ($(k_\mathrm{F}a_\mathrm{eff})^{-1}\lesssim 0$) is already seen in the mean-field level (see Fig.~\ref{fig:gap_v_ng} (b))~\cite{PhysRevA.82.063628,epjb.83.445}.
 Thus, it is simply due to the Fermi-surface mismatch between the $\uparrow$-spin and $\downarrow$-spin components.

 In the strong-coupling side ($(k_\mathrm{F}a_\mathrm{eff})^{-1}\gtrsim 0$), on the other hand, 
 SCTMA always gives $T_\mathrm{c}=0$ in the 2D-3D case ($t_\uparrow=0$), 
 which is in contrast to the mean-field result ($T_\mathrm{c}>0$) shown in Fig.~\ref{fig:gap_v_ng} (b).
 Since the latter indicates that the pairing interaction itself is strong enough to overcome the depairing by the Fermi surface mismatch, this SCTMA result is considered to reflect a strong-coupling effect in a 2D-3D Fermi-Fermi mixture, 
 which is completely ignored in the mean-field BCS theory.

 To confirm this, 
 it is helpful to evaluate the self-energy $\Sigma_\sigma(\bm{p},i\omega_m)$ in Eq.~\eqref{eq:Self_energy} within the so-called pseudogap approximation~\cite{CHEN20051,PhysRevB.66.024510,PhysRevA.80.033613}.
 That is, noting that the particle-particle scattering matrix $\Gamma(\bm{q},i\nu_n)$ (which describes pairing fluctuations) is strongly enhanced around $\bm{q}=\nu_n=0$ near $T_\mathrm{c}$, 
 we approximate Eq.~\eqref{eq:Self_energy} to 
 \begin{align}
    \Sigma_\sigma(\bm{p},i\omega_m)&\approx TG_{-\sigma}(\bm{-p},-i\omega_m)\sum_{\bm{q}, i\nu_n}\Gamma(\bm{q},i\nu_n) \notag\\
    &\approx TG_{-\sigma}(\bm{-p},-i\omega_m)\sum_{\bm{q}}\Gamma(\bm{q},0). \label{eq:selfenergy_pg}
 \end{align}
 Here, we have retained the most dominant term in the $\nu_n$-summation in the second line.
 In addition, expanding the denominator of Eq.~\eqref{eq:Scattering_matrix} at $\nu_n=0$ with respect to $\bm{q}=(\bm{q}_\perp,q_z)$ up to the second order, 
 one has
  \begin{align}
    \Sigma_\sigma(\bm{p},i\omega_m)\approx T&G_{-\sigma}(\bm{-p},-i\omega_m)\notag \\
    &\times\sum_{\bm{q}} \frac{-U}{[1-U\Pi(\bm{0},0)]+\alpha_\perp q_\perp^2+\alpha_z q_z^2}, \label{eq:3_3}
  \end{align}
  where
  \begin{align}
     \alpha_\perp&=-U\left.\frac{\partial^2\Pi(\bm{q},i\nu_n=0)}{\partial q_\perp^2}\right|_{\bm{q}=\bm{0}}, \\
     \alpha_z&=\left.-U\frac{\partial^2\Pi(\bm{q},i\nu_n=0)}{\partial q_z^2}\right|_{\bm{q}=\bm{0}}. \label{eq:alpha_z}
 \end{align}
 Particularly at $T_\mathrm{c}$, 
 since the Thouless criterion in Eq.~\eqref{eq:thouless_criterion} is satisfied, 
 Eq.~\eqref{eq:3_3} is reduced to
 \begin{align}
     \Sigma_\sigma(\bm{p},i\omega_m)\approx &TG_{-\sigma}(\bm{-p},-i\omega_m) \notag\\
     &\times\int_{-\pi/d_z}^{\pi/d_z}\frac{\dd{q_z}}{2\pi}\int_0^{q_\mathrm{c}}\frac{\dd{q_\perp}2\pi q_\perp}{(2\pi)^2}\frac{-U}{\alpha_\perp q_\perp^2+\alpha_z q_z^2}, \label{eq:3_4}
 \end{align}
 where $q_\mathrm{c}$ is a cutoff momentum.
 While Eq.~\eqref{eq:3_4} well converges in the 3D-3D case because of $\alpha_\perp>0$ and $\alpha_z>0$, 
 it exhibits a logarithmic infrared divergence in the 2D-3D case (except at $T=0$), 
 because $\alpha_z$ vanishes as shown in Appendix \ref{sec:tup0limit}.
 This immediately concludes that the Thouless criterion is never satisfied in the 2D-3D case when $T>0$.
 \begin{figure}[t]
   \centering
   \includegraphics[width=\linewidth]{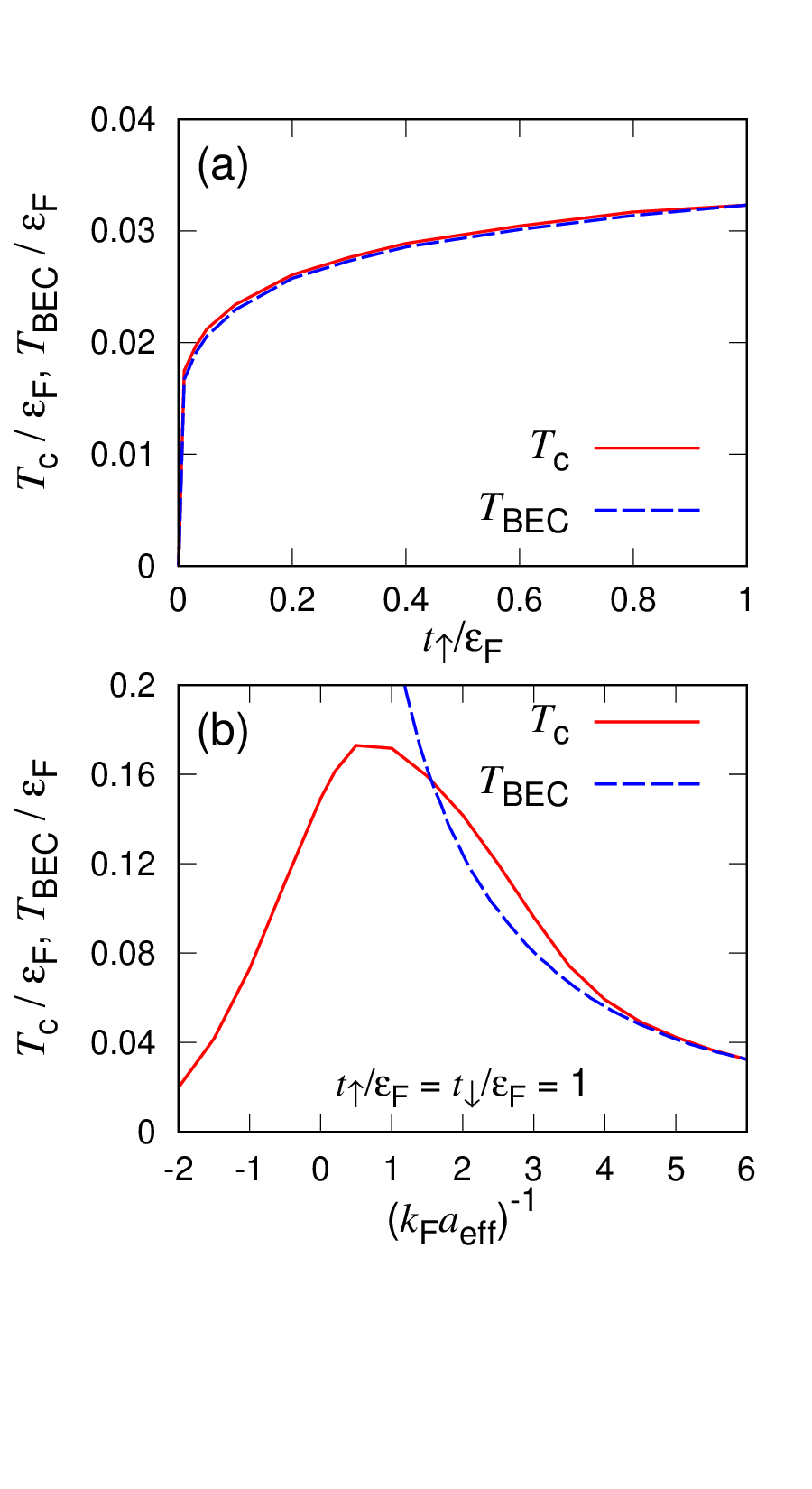}
   \vspace*{-3cm}
   \caption{
     Comparison of the calculated $T_\mathrm{c}$ in SCTMA with $T_\mathrm{BEC}$ obtained from Eq.~\eqref{eq:number_bose}.
     (a) $T_\mathrm{c}$ and $T_\mathrm{BEC}$ as functions of $t_\uparrow/\epsilon_\mathrm{F}$ in the strong-coupling regime when $(k_\mathrm{F}a_\mathrm{eff})^{-1}=6$.
     (b) Interaction dependence in the 3D-3D case ($t_\uparrow/\epsilon_\mathrm{F}=t_\downarrow/\epsilon_\mathrm{F}=1$).
   }
   \label{fig:Tc_balance}
 \end{figure}

 In the field of Fermi superfluids, 
 it is known that the BCS-type superfluid long-range order in 1D and 2D systems is prohibited at non-zero temperature due to strong low-dimensional pairing fluctuations, 
 which is also referred to as the Hohenberg-Mermin-Wagner theorem in the literature~\cite{PhysRevLett.17.1133,PhysRev.158.383}.
 In the present 2D-3D mixed-dimensional case, 
 Eq.~\eqref{eq:3_4} with $\alpha_z=0$ indicates that the suppression of $T_\mathrm{c}$ down to zero seen in the strong-coupling regime is caused by two-dimensional (in-plane) pairing fluctuations.
 Thus, the origin of this phenomenon is essentially the same as that in the two-dimensional case.

 In order to explain the reason why the three-dimensionality of the $\downarrow$-spin component does not work for the stabilization of the superfluid long-range order, 
 we consider the motion of Cooper pairs: 
 Deep inside the strong coupling BEC regime, where most Fermi atoms form $N_B=N/2$ Bose molecules, 
 system properties are well described by a gas of those molecules, 
 having the following single-particle dispersion (The derivation is summarized in Appendix \ref{sec:BEC_limit_calculation}.): 
 \begin{align}
     \epsilon_{\mathrm{B},\bm{q}}=\frac{q_\perp^2}{4m}+t_{\mathrm{B},z}[1-\cos(q_zd_z)], \label{eq:energy_boson}
 \end{align}
 where 
 \begin{align}
    t_{\mathrm{B},z}=\frac{2t_\uparrow t_\downarrow d_z^2}{|E_\mathrm{b}|}, \label{eq:t_B}
 \end{align} 
 describes the molecular hopping in the $z$-direction, and 
 \begin{align}
    E_\mathrm{b}=2(t_\uparrow+t_\downarrow)\big[1-\cosh[(a_\mathrm{eff}/d_z)^{-1}]\big]<0, \label{eq:E_b}
 \end{align}
 is the binding energy of a two-body bound state.
 Indeed, when one evaluates the BEC phase transition temperature $T_\mathrm{BEC}$ of this ideal Bose gas from the number equation, 
 \begin{align}
     N_\mathrm{B}=\sum_{\bm{q}}n_\mathrm{B}(\epsilon_{\mathrm{B},\bm{q}}), \label{eq:number_bose}
 \end{align}
 (where $n_\mathrm{B}(x)$ is the Bose distribution function)
 the calculated $T_\mathrm{BEC}$ agrees well with the SCTMA result, 
 as shown in Fig.~\ref{fig:Tc_balance}(a).

 Because a Cooper pair consists of $\uparrow$- and $\downarrow$-spin atoms, 
 the molecular hopping $t_\mathrm{B,z}$ in the $z$-direction is proportional to $t_\uparrow\times t_\downarrow$ in Eq.~\eqref{eq:t_B}.
 As a result, in a 2D-3D mixture ($t_\uparrow=0$ and $t_\downarrow\neq 0$), Cooper-pair molecules form a two-dimensional Bose gas, giving $T_\mathrm{BEC}=0$.
 When we view this from the anisotropy of the effective mass $\bm{m}_\mathrm{B}=(m_{\mathrm{B},x},m_{\mathrm{B},y},m_{\mathrm{B},z})$ of a Cooper-pair boson around the bottom of the molecular band in Eq.~\eqref{eq:energy_boson},
 one sees 
 \begin{align}
   \bm{m}_\mathrm{B}=\left(2m,2m,\frac{|E_\mathrm{b}|}{2t_\uparrow t_\downarrow d_z^2}\right).
 \end{align}
 Then, one may understand the vanishing $T_\mathrm{c}$ in the 2D-3D limit ($t_\uparrow=0$) as a result of the divergence of the effective mass $m_{\mathrm{B},z}$ in the $z$-direction.

 We note that Eq.~\eqref{eq:energy_boson} is also useful in understanding the strong-coupling behavior of $T_\mathrm{c}$ in a 3D-3D mixture ($t_\uparrow=t_\downarrow\neq 0$).
 In this case, although $t_\uparrow\times t_\downarrow$ in Eq.~\eqref{eq:t_B} is non-zero, 
 the molecular hopping $t_{B,z}$ in Eq.~\eqref{eq:t_B} again decreases to eventually vanish in the strong-coupling limit ($(a_\mathrm{eff}/d_z)^{-1}=(k_\mathrm{F}a_\mathrm{eff})^{-1}\rightarrow\infty$), 
 due to the increase of the binding energy $|E_\mathrm{b}|$ in Eq.~\eqref{eq:E_b}.
 As a result, $T_\mathrm{BEC}$ decreases with increasing interaction strength, to approach zero in the strong-coupling limit. 
 Indeed, this behavior well describes the interaction dependence of $T_\mathrm{c}$ in the strong-coupling regime, as shown in Fig.~\ref{fig:Tc_balance}(b). 
 \begin{figure}[t]
    \centering
    \includegraphics[width=\linewidth]{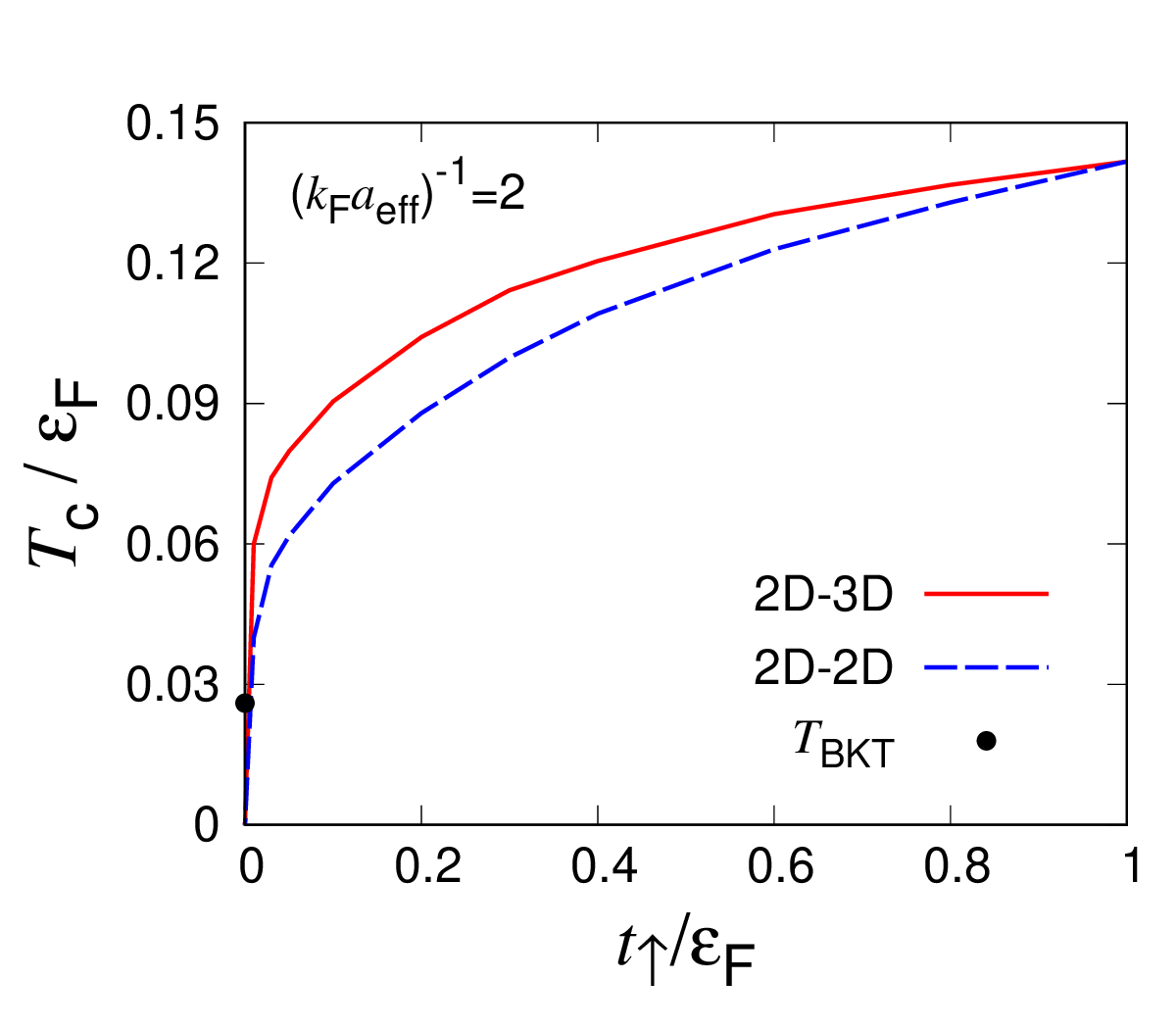}
    \vspace*{-0.5cm}
    \caption{
        Difference between 2D-2D and 2D-3D mixtures in the case when the inter-layer hopping $t_\sigma$ is introduced.
        In the 2D-3D case, only $t_\uparrow$ is increased with $t_\downarrow/\epsilon_\mathrm{F}=1$.
        In the 2D-2D case, the both $t_\uparrow$ and $t_\downarrow$ are increased, keeping $t_\uparrow=t_\downarrow$.
        In this figure, we also plot $T_{\mathrm{BKT}}$ in Eq.~\eqref{eq:T_BKT}.
    }
    \label{fig:Tc_MD_Q2D}
 \end{figure}

 From the effective molecular band in Eq.~\eqref{eq:energy_boson}, we find the difference between the 2D-3D system and the 2D-2D one (where $t_\downarrow$ also vanishes):
 When we start from these systems to approach the 3D-3D mixture, 
 while the inter-layer hopping $t_{\mathrm{B},z}$ in Eq.~\eqref{eq:t_B} linearly increases with $t_\uparrow$ in the 2D-3D case ($t_{\mathrm{B},z}\propto t_\uparrow$), 
 in the 2D-2D case it increases quadratically ($t_{\mathrm{B},z}\propto t_\uparrow^2$) 
 as one increases $t_\uparrow$ and $t_\downarrow$ keeping $t_\uparrow=t_\downarrow$.
 As a result, the recovery of $T_\mathrm{c}$ with increasing $t_\uparrow$ is faster in the 2D-3D case than the 2D-2D case, 
 as seen in Fig.~\ref{fig:Tc_MD_Q2D}.

 Regarding 2D Fermi superfluids at finite temperatures, 
 although the BCS-type long-range order is prohibited, the Berezinskii-Kosterlitz-Thouless (BKT) transition is possible.
 Simply applying the previous work on the BKT phase transition temperature $T_\mathrm{BKT}$ in a 2D Fermi gas~\cite{PhysRevLett.103.165301} to the strong-coupling regime of the present 2D-3D mixture, one finds
 \begin{align}
    T_\mathrm{BKT}=\frac{1}{8}\epsilon_{\mathrm{F}}^\mathrm{2D}, \label{eq:T_BKT}
 \end{align}
 where $\epsilon_{\mathrm{F}}^\mathrm{2D}$ is the Fermi energy of a 2D Fermi gas~\cite{PhysRevLett.103.165301}, 
 which is related to $\epsilon_\mathrm{F}$ in the present 2D-3D mixture as $\epsilon_{\mathrm{F}}^\mathrm{2D}=(2/3\pi)\epsilon_{\mathrm{F}}$.

 In Fig.~\ref{fig:Tc_MD_Q2D}, we compare $T_\mathrm{BKT}$ in Eq.~\eqref{eq:T_BKT} with the BCS-type superfluid transition temperature $T_\mathrm{c}$.
 Although it remains as an interesting future problem how the BKT phase transition changes to the BCS one as one goes away from the 2D-3D limit, 
 because $T_\mathrm{c}$ is found to rapidly restore around $t_\uparrow=0$, 
 the most region of the (2D-2D)-(2D-3D) crossover is dominated by the BCS-type long-range order.

 We briefly note that the increase of $T_\mathrm{c}$ around $t_\uparrow=0$ seen in Figs.~\ref{fig:Tc_balance}(a) and \ref{fig:Tc_MD_Q2D} can be analytically understood deep inside the strong-coupling regime.
 The result is (For the derivation, see Appendix~\ref{sec:BEC_limit_calculation}.), 
  \begin{align}
  T_\mathrm{c}&\approx \frac{\pi N}{2md_z}\left[
      W\left(\frac{\pi N|E_\mathrm{b}|}{4md_z t_\uparrow t_\downarrow}\right)
  \right]^{-1} \; \; \; \; \; \; \; \; \;  (t_\uparrow\ll 1)\notag\\
  &\approx \frac{\pi N}{2md_z}\left[\ln\left(\frac{\pi N|E_\mathrm{b}|}{4md_z t_\uparrow t_\downarrow}\right)-\ln\left[\ln\left(\frac{\pi N|E_\mathrm{b}|}{4md_z t_\uparrow t_\downarrow}\right)\right]\right]^{-1}, \label{eq:T_BEC}
 \end{align}
 where $W(Y)$ is the Lambert W function, which satisfies $Y=Xe^X, X=W(Y)$.
 We then find that the behavior of $T_\mathrm{c}$ around $t_\uparrow=0$ comes from the inverse of a logarithmic divergence with respect to $t_\uparrow$.
 \begin{figure}[t]
    \centering
    \includegraphics[width=\linewidth]{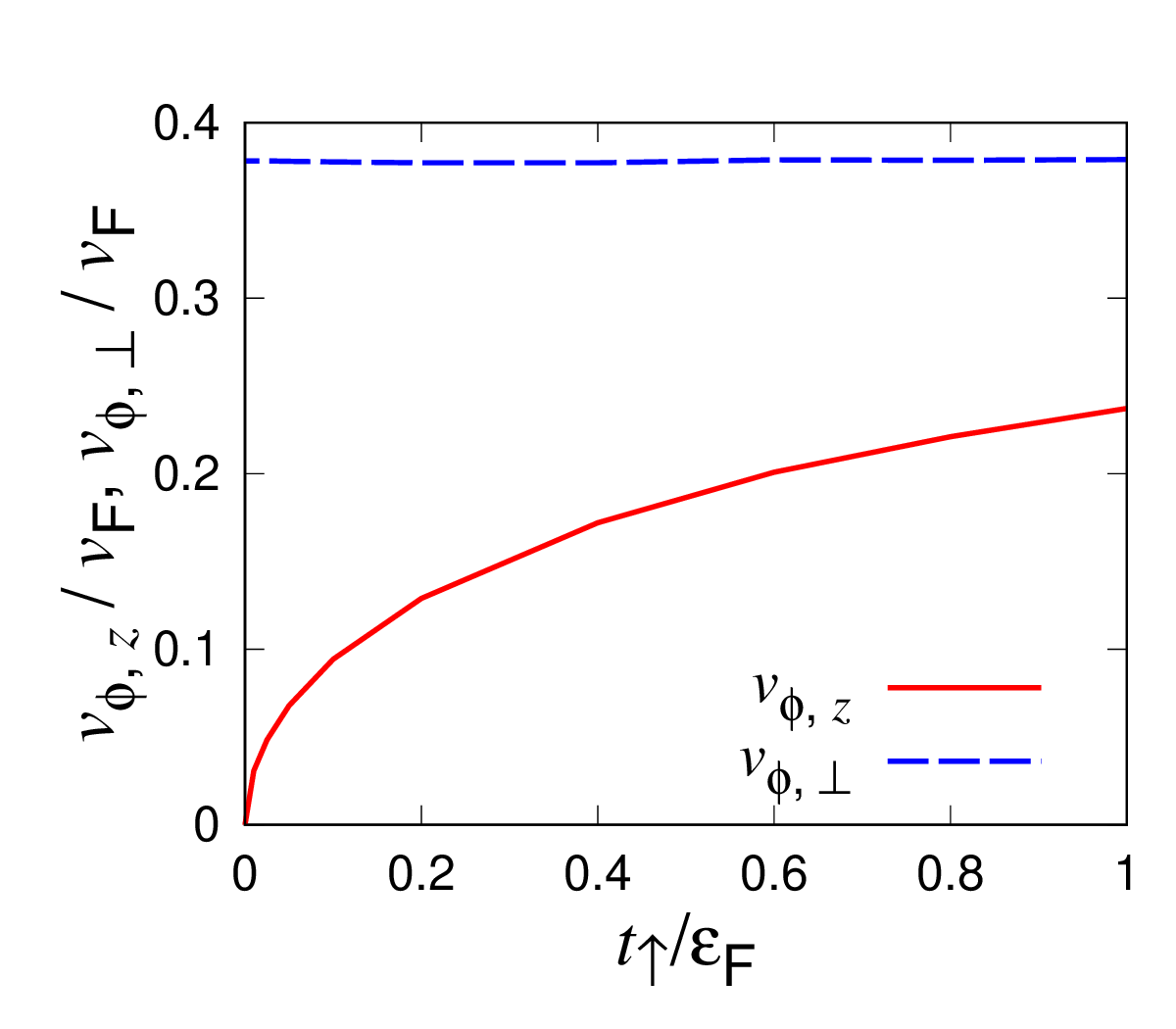}
    \caption{
        Sound velocity $\bm{v}=(\bm{v}_{\phi,\perp},v_{\phi,z})$ of the Goldstone mode at $T=0$, 
        calculated in the coupled mean-field BCS theory with GRPA.
        We set $(k_\mathrm{F}a_\mathrm{eff})^{-1}=2$.
    }
    \label{fig:v_ng_2D3D}
  \end{figure}

 Before ending this section, 
 we point out that, 
 although the mean-field BCS theory cannot capture the two-dimensional aspect of a 2D-3D mixture as far as we consider $T_\mathrm{c}$ and $\Delta$ in Fig.~\ref{fig:gap_v_ng}, 
 the mean-field scheme is still valid for this purpose in the analysis of the Goldstone mode: 
 Figure~\ref{fig:v_ng_2D3D}  shows the sound velocity $\bm{v}=(\bm{v}_{\phi,\perp},v_{\phi,z})$ of the superfluid Goldstone mode at $T=0$, 
 being calculated within the framework of the coupled mean-field BCS theory with the generalized random phase approximation (GRPA), 
 where the pair-correlation function is evaluated in RPA.
 (We explain the outline of this scheme in Appendix~\ref{sec:GRPA}.)
 We see in this figure that, 
 while the propagation of the Goldstone mode in the 2D in-plane direction is still possible ($\bm{v}_{\phi,\perp}\neq \bm{0}$) in a 2D-3D mixture, 
 $v_{\phi,z}$ vanishes when $t_\uparrow=0$.
 Since the Goldstone mode is a collective oscillation of the superfluid order parameter, 
 it physically means the absence of the phase coherence in the $z$-direction in the 2D-3D case.

 \section{Single-particle excitations and three-dimensionality in 2D-3D mixtures}\label{sec:DOS}
\begin{figure}[t]
  \centering
  \includegraphics[width=\linewidth]{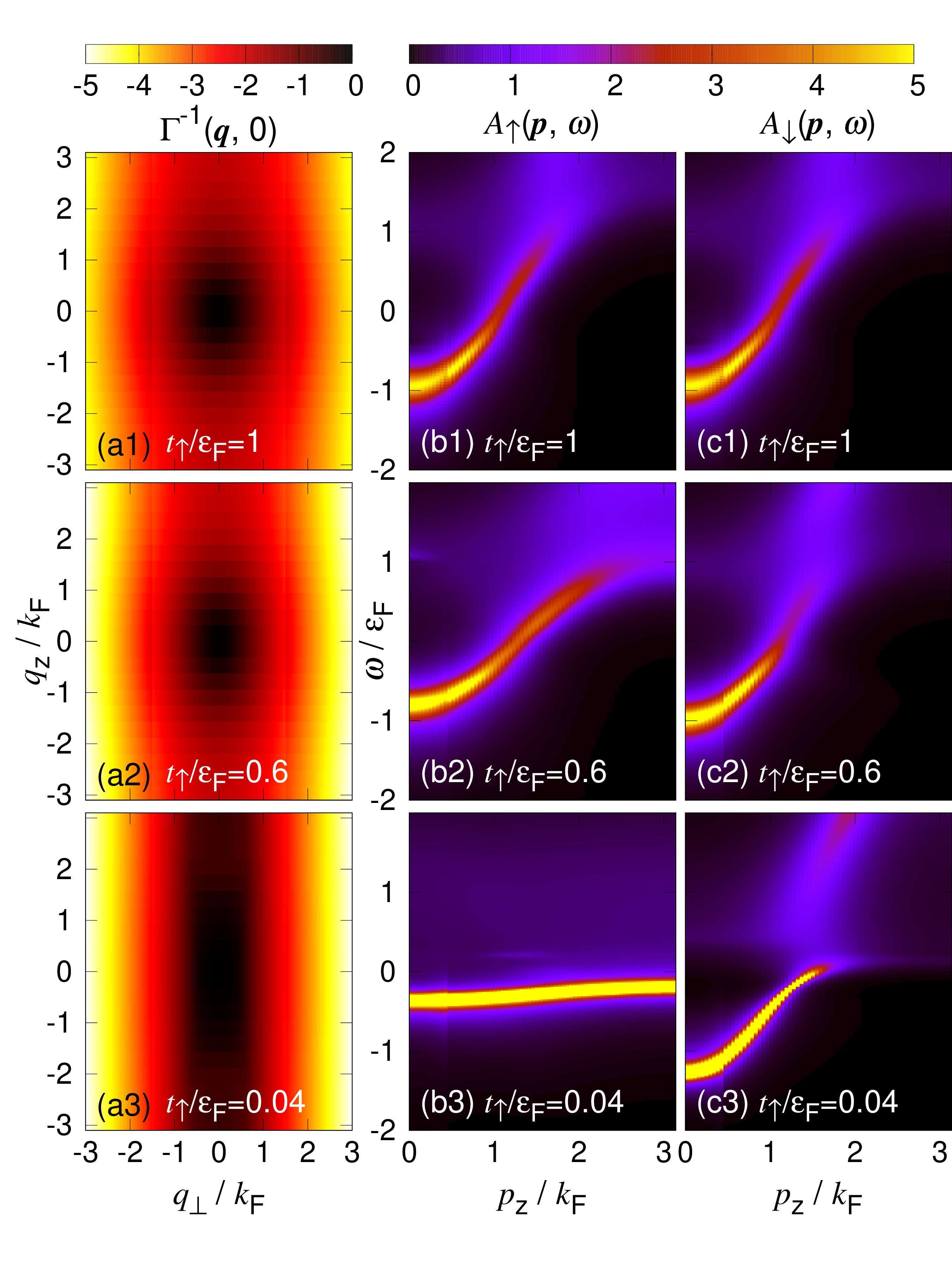}
  \caption{
  (a1)-(a3) Intensity of $\Gamma^{-1}(q_\perp, q_z, 0)$ in the $q_\perp$-$q_z$ plane. 
  (b1)-(b3) ((c1)-(c3)) Intensity of single-particle spectral weight $A_\uparrow(\bm{p},\omega)$($A_\downarrow(\bm{p},\omega)$) in the $p_z$-$\omega$ plane.
  We set $T=T_\mathrm{c}$ and $(k_\mathrm{F}a_\mathrm{eff})^{-1}=0$.
  For $A_{\sigma=\uparrow,\downarrow}(\bm{p},\omega)$, we take $p_\perp/k_\mathrm{F}=0.03$.
  Since $\Gamma(\bm{q},0)$ divergence at $\bm{q}=\bm{0}$, we plot $\Gamma^{-1}(\bm{q},0)$ in panel (a1)-(a3). 
  }
  \label{fig:gamma_spec}
 \end{figure}

 The left panels in Fig.~\ref{fig:gamma_spec} show the structure of the particle-particle scattering matrix $\Gamma(\bm{q},i\nu_n=0)$ in momentum space.
 As expected from the discussions in Sec.~\ref{sec:Tc}, 
 it gradually loses the $q_z$-dependence with decreasing the value of $t_\uparrow$ (see Fig.~\ref{fig:gamma_spec}(a1)-(a3)).
 Of course, this $t_\uparrow$-dependence is consistent with that of the $\uparrow$-spin component $A_\uparrow(\bm{p},\omega)$ of the single-particle spectral weight shown in Figs.~\ref{fig:gamma_spec}(b1)-(b3), 
 where the peak line, 
 describing the $\uparrow$-spin component of single-particle dispersion in the $p_z$-direction, 
 gradually becomes flat as one approaches the 2D-3D limit.

 However, although the motion of $\downarrow$-spin atoms is restricted to the 2D in-plane direction when they form Cooper pairs in a 2D-3D mixture, we find from Figs.~\ref{fig:gamma_spec}(c1)-(c3) that single-particle dispersion still possesses three-dimensionality, 
 irrespective of the value of $t_\uparrow$.
 This means that, once $\downarrow$-spin atoms are released from Cooper pairs, they re-acquire the three-dimensional motion.

 In Fig.~\ref{fig:gamma_spec}(c3), one sees a gap-line structure around $p_z/k_\mathrm{F}\cong 1.5$.
 Since the superfluid order parameter $\Delta$ vanishes at $T_\mathrm{c}$, 
 it is a pseudogap associated with pairing fluctuations,
 that are enhanced by the two-dimensional properties of a nearly 2D-3D mixture.
 In contrast, such a strong-coupling correction is not seen in Fig.~\ref{fig:gamma_spec}(b3).

 To grasp the origin of this difference between Figs.~\ref{fig:gamma_spec}(b3) and (c3), 
 the pseudogap approximation given in Eq.~\eqref{eq:selfenergy_pg} is again useful:
 In the case of 2D-3D mixtures, noting that $\Gamma(\bm{q},i\nu_n)$ ($G_\uparrow(\bm{p},i\omega_m)$) does not depend on $q_z$ ($p_z$), 
 $\Sigma_\downarrow(\bm{p},i\omega_m)$ is approximated to 
 \begin{align}
    \Sigma_\downarrow(\bm{p},i\omega_m)\approx -\Delta_{\mathrm{PG}}^2G_{0,\uparrow}(-\bm{p}_\perp,-i\omega_m), \label{eq:SE_pg_dw}
 \end{align}
 where we have replaced the dressed Green's function $G_\uparrow(\bm{p}_\perp,i\omega_m)$ by the bare one, 
 $G_{0,\uparrow}(\bm{p}_\perp,-i\omega_m)=[-i\omega_m-\xi_{\bm{p}_\perp,\uparrow}]^{-1}$, for simplicity.
 In Eq.~\eqref{eq:SE_pg_dw}, 
 \begin{align}
    \Delta_{\mathrm{PG}}^2=-\sum_{\bm{q}_\perp}\Gamma(\bm{q}_\perp,0)
 \end{align}
 is the pseudogap parameter, 
 describing the strength of pairing fluctuations.
 Substituting Eq.~\eqref{eq:SE_pg_dw} into $G_\downarrow(\bm{p},i\omega_m)$ in Eq.~\eqref{eq:Greens_func}, 
 we have, after carrying out the analytic continuation, $i\omega_m\rightarrow \omega_+\equiv\omega+i\delta$,
 \begin{align}
    G_\downarrow(\bm{p}, \omega_+) &= \frac{1}{\omega_+-\xi_{\bm{p},\downarrow}-\frac{\Delta_{\mathrm{PG}}^2}{\omega_+ +\xi_{\bm{p}_\perp,\uparrow}}}. \label{eq:Greensfunction_PG_dw}
 \end{align}
 Equation~\eqref{eq:Greensfunction_PG_dw} clearly shows that the 3D $\downarrow$-spin particle band $\omega=\xi_{\bm{p},\downarrow}$ couples with the 2D $\uparrow$-spin hole band $\omega=-\xi_{\bm{p}_\perp,\uparrow}$ with the coupling strength $\Delta_{\mathrm{PG}}^2$.
 When these bands cross each other, the level repulsion occurs between them, 
 which gives the (pseudo)gap.
 Looking for the poles of Eq.~\eqref{eq:Greensfunction_PG_dw}, 
 we obtain the gapped two bands with the energy gap $2\Delta_{\mathrm{PG}}$, 
 given by 
 \begin{align}
    \omega_{\pm} &= \frac{\xi_{\bm{p},\downarrow}-\xi_{\bm{p}_\perp,\uparrow}}{2}
    \pm \sqrt{\left(\frac{\xi_{\bm{p},\downarrow}+\xi_{\bm{p}_\perp,\uparrow}}{2}\right)^2+\Delta_{\mathrm{PG}}^2}, \label{eq:375}
 \end{align}
 which qualitatively explains the pseudogapped peak line seen in Fig.~\ref{fig:gamma_spec}(c3).

 Applying the same analysis to the 2D $\uparrow$-spin component, 
 we obtain a bit different expression from Eq.~\eqref{eq:Greensfunction_PG_dw} as
 \begin{align}
    G_\uparrow(\bm{p}, \omega_+) &= \frac{1}{\omega_+-\xi_{\bm{p}_\perp,\uparrow}-\Delta_{\mathrm{PG}}^2\sum_{p_z}\frac{1}{\omega_+ +\xi_{\bm{p},\downarrow}}}. \label{eq:Greensfunction_PG_up}
 \end{align}
 We briefly note that the $p_z$-summation comes from the $q_z$-independence of $\Gamma(\bm{q},i\nu_n)$.
 Because of this $p_z$-summation, the $\uparrow$-spin band does not reflect the three dimensionality of the $\downarrow$-spin band, 
 which explains the (almost) $p_z$-independent peak line seen in Fig.~\ref{fig:gamma_spec}(b3).

 \begin{figure}[t]
    \centering
    \includegraphics[width=\linewidth]{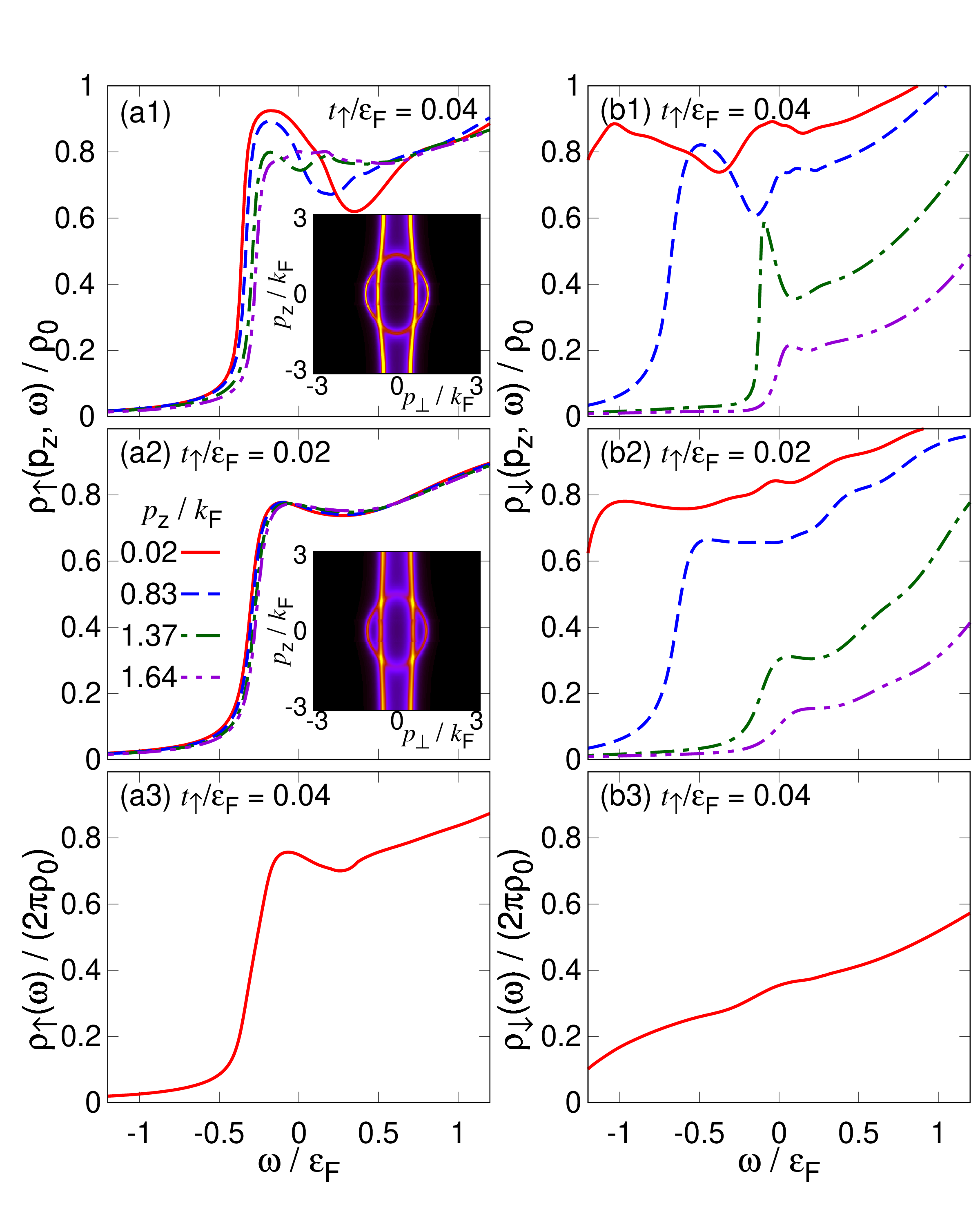}
    \caption{
      Calculated $p_z$-resolved DOS $\rho_\sigma(p_z,\omega)$ in Eq.~\eqref{eq:rho_pz} (top and middle panels) 
      and DOS $\rho_\sigma(\omega)$ in Eq.~\eqref{eq:rho_omega} (bottom panels).
      We take $(k_\mathrm{F}a_{\mathrm{eff}})^{-1}=0$, $t_\uparrow/\epsilon_\mathrm{F}=0.04$ at $T_\mathrm{c}$ in the top and bottom panels, 
      and $t_\uparrow/\epsilon_\mathrm{F}=0.02$ at $T/\epsilon_\mathrm{F}=0.05$ in the middle panels.
      The insets in (a1) and (a2) show $A_\uparrow(\bm{p}, \omega=0) + A_\downarrow(\bm{p}, \omega=0)$, where the peaks correspond to the Fermi surfaces.
      DOS is normalized by $\rho_0 = mk_\mathrm{F}/\pi$.
      }
    \label{fig:DOS_Uni}
 \end{figure}

 Figure~\ref{fig:DOS_Uni} shows the ($p_z$-resolved) DOS in Eqs.~\eqref{eq:rho_omega} and \eqref{eq:rho_pz} near the 2D-3D limit ($t_\uparrow/\epsilon_\mathrm{F}\ll 1$).
 For the (almost 2D) $\uparrow$-spin component, we expect from $G_\uparrow(\bm{p},\omega_+)$ in Eq.~\eqref{eq:Greensfunction_PG_up} that 
 single-particle excitations gradually lose the $p_z$-dependence.
 Indeed, while the $p_z$-dependent pseudogap (dip around $0\lesssim\omega/\epsilon_\mathrm{F}\lesssim 0.5$) is seen in the $p_z$-resolved DOS $\rho_\uparrow(p_z,\omega)$ when $t_\uparrow/\epsilon=0.04$ (panel (a1)), 
 it becomes almost $p_z$-independent when $t_\uparrow/\epsilon_\mathrm{F}=0.02$ (panel (a2)).

 On the other hand, 
 $G_\downarrow(\bm{p},\omega_+)$ in Eq.~\eqref{eq:Greensfunction_PG_dw} makes us expect that the $\downarrow$-spin component continues to possess three-dimensional single-particle properties and 
 to exhibit the pseudogap phenomenon (as far as the $\uparrow$-spin band crosses the $\downarrow$-spin band) 
 as one approaches the 2D-3D mixture with decreasing $t_\uparrow$.
 Indeed, we can confirm these expectations from the $p_z$- and $\omega$-dependence of $\rho_\downarrow(p_z,\omega)$ shown in Figs.~\ref{fig:DOS_Uni}(b1) and (b2).

 One sees in Fig.~\ref{fig:DOS_Uni}(b1) and (b2) that the pseudogap (dip) position moves to higher energies with increasing $p_z$.
 Regarding this, we note that, 
 in the pseudogap approximation, the central energy ($\equiv \omega_{\mathrm{PG}}$) of the pseudogap structure is given by the first term in Eq.~\eqref{eq:375}.
 When we simply extend Eq.~\eqref{eq:375} to the case of ``nearly'' 2D-3D mixture, 
 we again obtain Eq.~\eqref{eq:375} where $\bm{p}_\perp$ in $\xi_{\bm{p}_\perp,\uparrow}$ is replaced by $\bm{p}$.
 Then, we have 
 \begin{align}
    \omega_{\mathrm{PG}}&=\frac{\xi_{\bm{p},\downarrow}-\xi_{\bm{p},\uparrow}}{2} \notag \\
    &=(t_\uparrow-t_\downarrow)[1-\cos(p_zd_z)]-(\mu_\downarrow-\mu_\uparrow), \label{eq:omega_center}
 \end{align}
 which increases with increasing $p_z$ as far as $t_\downarrow>t_\uparrow$.
 Thus the $p_z$-dependence of the pseudogap position seen in Figs.~\ref{fig:DOS_Uni}(b1) and (b2) comes from the difference between almost 2D $\uparrow$-spin band and 3D $\downarrow$-spin one.

 We see in Fig.~\ref{fig:DOS_Uni}(a1) the opposite $p_z$-dependence of the pseudogap position.
 Regarding this, further extending Eq.~\eqref{eq:375} to the $\uparrow$-spin component, 
 we obtain Eq.~\eqref{eq:omega_center} where ``$\uparrow$'' and ``$\downarrow$'' are exchanged, 
 which immediately explains the decrease of the pseudogap position with increasing $p_z$ seen in Fig.~\ref{fig:DOS_Uni}(a1).

 As a common tendency about the pseudogap phenomenon seen in Figs.~\ref{fig:DOS_Uni}(a1), (a2), (b1) and (b2), 
 the dip structure becomes obscure when $p_z$ is large.
 To explain the reason for this, we show in the insets in Figs~\ref{fig:DOS_Uni}(a1) and (a2), the Fermi surfaces of the $\uparrow$-spin (cylindrical shape) and $\downarrow$-spin (spherical shape) components determined from the single-particle spectral weight.
 Since the coupling between the particle band $\omega=\xi_{\bm{p},\sigma}$ and the hole band $\omega=-\xi_{-\bm{p},-\sigma}$ by pairing fluctuations is the essence of the pseudogap phenomenon, 
 these insets imply that this phenomenon in $\rho_\sigma(p_z,\omega)$ becomes obscure when $p_z/k_\mathrm{F}\gtrsim 1.6$, 
 where the $\downarrow$-spin surface does not exist.
 Indeed, the dip structure commonly disappears above this value in Figs.~\ref{fig:DOS_Uni}(a1), (b1) and (b2).

 Before ending this section, 
 we briefly note that, 
 because the pseudogap position depends on $p_z$ (see Figs.~\ref{fig:DOS_Uni}(a1) and (b1)), 
 the pseudogap is not so clearly seen in DOS $\rho_\sigma(\omega)$ given by summing $\rho_\sigma(p_z,\omega)$ over $p_z$ (panels (a3) and (b3)).
 Thus, to observe this many-body phenomenon in a mixed-dimensional Fermi gas, 
 especially near the 2D-3D limit, 
 the single-particle spectral weight $A_\sigma(\bm{p},\omega)$ would be more effective.

 \section{Summary}\label{sec:summary}

 To summarize, we have theoretically investigated a strongly interacting mixed-dimensional two-component Fermi atomic gas.
 Including mixed-dimensional pairing fluctuations within the framework of SCTMA, 
 we examined how the superfluid instability is affected by dimensional imbalance between Fermi atoms contributing to the Cooper-pair formation.
 
 We showed that pairing fluctuations are dominated by lower dimensional component, because the motion of a Cooper pair is restricted by this component.
 As a result, in a 2D-3D Fermi-Fermi mixture, 
 the BCS-type superfluid long-range order is completely destroyed when $T>0$, 
 by two-dimensional pairing fluctuations.
 Thus, while the long-range order is supported by the three-dimensionality of the system in an ordinary 3D Fermi superfluid, 
 the 3D component in a 2D-3D mixture does not play this role at all.
 In this sense, the background physics of the vanishing $T_\mathrm{c}$ by 2D-3D pairing fluctuations is essentially the same as the Hohenberg-Mermin-Wagner theorem discussed in low-dimensional Fermi superfluids.

 Although the two-dimensionality of a 2D-3D mixture originates from pairing fluctuations, 
 which are completely ignored in the mean-field BCS theory, 
 we pointed out that the mean-field theory can still capture this to some extent when we consider the Goldstone mode.
 Indeed, the calculated mode velocity by the coupled mean-field BCS theory with GRPA exhibits two-dimensional character in a 2D-3D mixture, 
 although the BCS theory itself gives non-zero $T_\mathrm{c}$ there.

 While the superfluid instability is dominated by the two-dimensionality of a 2D-3D mixture, 
 we clarified that the three-dimensionality of this system plays important roles for the pseudogap phenomenon.
 That is, the existence of the 3D component was shown to be crucial for the appearance of the pesudogapped single-particle spectral weight, 
 as well as the dip position in the $p_z$-resolved DOS.
 In this sense, the dimensionality of a 2D-3D Fermi-Fermi mixture depends on what we observe.

 In this paper, we have only considered the BCS-type superfluid.
 Regarding this, since system properties are dominated by two-dimensional pairing fluctuations in a 2D-3D mixture, 
 we expect the BKT phase transition there.
 In this sense, it is an interesting future problem to examine how the 3D component affects this 2D quasi-long range order.
 In addition, it also remains to be solved how this BKT state changes to the BCS state as the system changes from the 2D-3D to the 3D-3D cases. 
 Since the realization of dimensional imbalance is not so easy in metallic superconductivity, 
 our results would contribute to the study of this unique many-body system by using highly tunable ultracold Fermi atomic gases.

 \par

 \par

 \par
 \begin{acknowledgments}
    We thank S. Uchida, Y. Okajima, K. Takemoto, H. Tsukada, H. Yamada, T. Kawamura and R. Hanai  for discussions.
    H.T. was supported by a Grant-in-Aid for JST SPRING (Grant No.JPMJSP2123).
    Y.O. was supported by a Grant-in-Aid for Scientific Research from MEXT and JSPS (Grants No.22KO3486 and No.26KO6983).
 \end{acknowledgments}
 \par
 \appendix
 \par

 \section{Mean-field BCS theory for a mixed-dimensional Fermi gas}\label{sec:mf_BCS}
 We explain the outline of the mean-field treatment of the model mixed-dimensional Hamiltonian in Eq.~\eqref{eq:Hamiltonian}~\cite{PhysRevA.82.063628}: 
 In two-component Nambu representation~\cite{doi:10.1143/JPSJ.66.2437}, 
 Eq.~\eqref{eq:Hamiltonian} can be written as 
 \begin{align}
     H 
     =&\frac{|\Delta|^2}{U}+\sum_{\bm{p}}\hat{\Psi}_{\bm{p}}^\dagger(\xi_{+,\bm{p}}\tau_3+\xi_{-,\bm{p}}\tau_0-\Delta\tau_1)\hat{\Psi}_{\bm{p}} \notag \\
     &+\sum_{\bm{p}}\xi_{\bm{-p},\downarrow}-\frac{U}{4}\sum_{\bm{q}, i=1,2}\rho_{i,\bm{q}}\rho_{i,-\bm{q}}, \label{eq:Ap_Hamiltonian}
 \end{align}
 where $\hat{\Psi}_{\bm{p}}^\dagger\equiv(c_{\bm{p},\uparrow}^\dagger, c_{-\bm{p},\downarrow})$ is the Nambu field and the Pauli matrices $\tau_{i=1,2,3}$ act on the particle-hole space.
 In Eq.~\eqref{eq:Ap_Hamiltonian}, 
 $\Delta = U\sum_{\bm{p}}\Braket{c_{\bm{-p},\downarrow}c_{\bm{p},\uparrow}}$ is the BCS superfluid order parameter, $\xi_{\pm,\bm{p}}=(\xi_{\bm{p},\uparrow}\pm\xi_{\bm{p},\downarrow})/2$, and 
 \begin{align}
    \rho_{i=1,2,\bm{q}}&\equiv\sum_{\bm{p}}\Psi_{\bm{p+q}/2}^\dagger\tau_i\Psi_{\bm{p-q}/2}\label{eq:rho_123},
 \end{align}
 are generalized density operators~\cite{doi:10.1143/JPSJ.66.2437}, 
 where $\rho_{1,\bm{q}}$ and $\rho_{2,\bm{q}}$ physically describe amplitude and phase fluctuations of the order parameter, respectively.
 \begin{figure}[t]
    \centering
    \includegraphics[width=\linewidth]{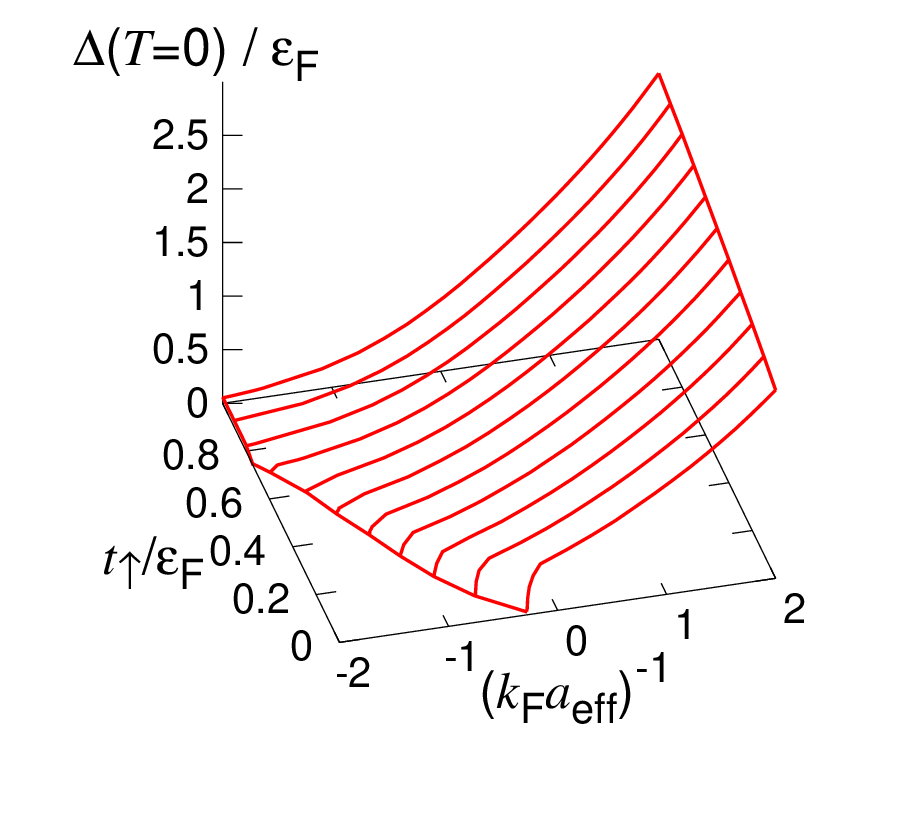}
    \vspace*{-1.5cm}
    \caption{
        Calculated BCS superfluid order parameter $\Delta(T=0)$ at $T=0$, as a function of the interaction strength $(k_\mathrm{F}a_\mathrm{eff})^{-1}$ and the dimensional imbalance tuned by $t_\uparrow$.
        (Note that $t_\downarrow/\epsilon_\mathrm{F}=1$.)
    }
    \label{fig:gap_splot}
 \end{figure}

 In the mean-field approximation, 
 the last term in Eq.~\eqref{eq:Ap_Hamiltonian} is ignored.
 The resulting Hamiltonian then has the bilinear form with respect to the Nambu field, 
 so that the system can simply be described by the $2\times 2$ matrix single-particle thermal Green's function, 
 having the form~\cite{doi:10.1143/JPSJ.66.2437}, 
 \begin{align}
    \hat{G}^0(\bm{p},i\omega_m) = \frac{1}{
        i\omega_m - (\xi_{+,\bm{p}}\tau_3+\xi_{-,\bm{p}}-\Delta\tau_1)
    }.\label{eq:G0}
 \end{align}
 The number equation for the $\uparrow$-spin ($\downarrow$-spin) component is then immediately obtained from the (11)-component ((22)-component) of $\hat{G}^0(\bm{p},i\omega_m)$ in Eq.~\eqref{eq:G0} as 
 \begin{align}
    N_\uparrow=\frac{1}{2}\sum_{\bm{p}}&\left[
        \left(1+\frac{\xi_{+,\bm{p}}}{E_{\bm{p}}}\right)n_{\mathrm{F}}(E_{\bm{p}}+\xi_{-,\bm{p}})\right.\notag \\
        &+\left.
        \left(1-\frac{\xi_{+,\bm{p}}}{E_{\bm{p}}}\right)n_{\mathrm{F}}(-E_{\bm{p}}-\xi_{-,\bm{p}})
    \right],\label{eq:0temp_num_up}\\
    N_\downarrow=\frac{1}{2}\sum_{\bm{p}}&\left[
        \left(1+\frac{\xi_{+,\bm{p}}}{E_{\bm{p}}}\right)n_{\mathrm{F}}(E_{\bm{p}}-\xi_{-,\bm{p}})\right.\notag \\
        &+\left.
        \left(1-\frac{\xi_{\bm{p},+}}{E_{\bm{p}}}\right)n_{\mathrm{F}}(-E_{\bm{p}}-\xi_{-,\bm{p}})
    \right],\label{eq:0temp_num_dw}
 \end{align}
 where $E_{\bm{p}}=\sqrt{\xi_{+,\bm{p}}^2+|\Delta|^2}$.

 The BCS gap equation is also obtained from the (12)-component of $\hat{G}^0(\bm{p},i\omega_m)$.
 The result is 
 \begin{align}
    1=-\frac{4\pi a_\mathrm{eff}}{m}\sum_{\bm{p}}&\left[
        \frac{1-n_{\mathrm{F}}(E_{\bm{p}}+\xi_{-,\bm{p}})-n_{\mathrm{F}}(E_{\bm{p}}-\xi_{-,\bm{p}})}{2E_{\bm{p}}} \right.\notag\\
        &-\left.
        \frac{1}{\epsilon_{\bm{p},\uparrow}+\epsilon_{\bm{p},\downarrow}}
    \right].\label{eq:gap}
 \end{align}
 Here, $n_{\mathrm{F}}(\epsilon)$ is the Fermi distribution function, and
 we have removed the ultraviolet divergence from the gap equation, 
 by using the effective $s$-wave scattering length $a_\mathrm{eff}$ in Eq.~\eqref{eq:a_eff}.

 Figure \ref{fig:gap_splot} shows the BCS-type mean-field superfluid order parameter $\Delta(T=0)$ at $T=0$, 
 being obtained from the coupled gap equation \eqref{eq:gap} with the number equations \eqref{eq:0temp_num_up} and \eqref{eq:0temp_num_dw}.
 (For detailed parameters, see the last paragraph in Sec.~\ref{sec:Formulation}.)
 Figure~\ref{fig:gap_v_ng}(a) shows the result which is extracted from Fig.~\ref{fig:gap_splot} at $t_\uparrow/\epsilon_\mathrm{F}=0$.
 In this 2D-3D limit, 
 solving these coupled equations at $T>0$ by setting $\Delta\rightarrow 0$, 
 we obtain the $T_\mathrm{c}$-line in Fig.~\ref{fig:gap_v_ng}(b).

 \section{Derivation of $\alpha_z=0$ in the 2D-3D limit} \label{sec:tup0limit}

  We show that $\alpha_z$ in Eq.~\eqref{eq:alpha_z} vanishes in the 2D-3D limit: 
  During the scattering process by the contact-type interaction involved in the model Hamiltonian in Eq.~\eqref{eq:Hamiltonian}, 
  because the annihilation of incident atoms and the creation of scattered ones occur at the same spatial position, 
  the pairing interaction does not induce any inter-layer hopping in the 2D $\uparrow$-spin component.
  As a result, the dressed SCTMA Green's function $G_\uparrow(\bm{p},i\omega_m)$ in Eq.~\eqref{eq:Greens_func} still has no $p_z$-dependence ($\equiv G_\uparrow(\bm{p}_\perp,i\omega_m)$), 
 as in the non-interacting case.
 Then, the pair-correlation function, 
 \begin{align}
    \Pi(\bm{q}, i\nu_n) = -T\sum_{\bm{p}, i\omega_m} G_\uparrow(\bm{p}_{\perp}, i\omega_m) G_\downarrow(\bm{q-p}, i\nu_n-i\omega_m),
 \end{align}
 in the 2D-3D case is found to be $q_z$-independent, 
 because $q_z$-dependence of the right hand side is removed by simply changing the variable as $q_z-p_z\rightarrow p_z$.
 This immediately concludes
 \begin{align}
    \alpha_z = -U\left.\frac{\partial^2\Pi(\bm{q},i\nu_n=0)}{\partial q_z^2}\right|_{\bm{q}=\bm{0}} = 0.
 \end{align}
 
 \section{Analytic results in the strong-coupling limit} \label{sec:BEC_limit_calculation}

 We provide the derivation of Eq.~\eqref{eq:energy_boson}-\eqref{eq:number_bose} and \eqref{eq:T_BEC}.
 In the strong-coupling limit, the pair corelation function in Eq.~\eqref{eq:PI} can be approximated to 
 \begin{align}
    \Pi(\bm{q},i\nu_n)=& 
    T\sum_{\bm{p},i\omega_m}G_{0,\uparrow}(\bm{p+q/2},i\nu_n+i\omega_m)\notag \\
      &\ \ \ \ \ \ \ \ \ \ \ \ 
      \times G_{0,\downarrow}(\bm{-p+q/2},-i\omega_m)\notag \\
    =&\sum_{\bm{p}}\frac{1-n_\mathrm{F}(\xi_{\bm{p}+\bm{q}/2,\uparrow})-n_\mathrm{F}(\xi_{-\bm{p}+\bm{q}/2,\downarrow})}{\xi_{\bm{p}+\bm{q}/2,\uparrow}+\xi_{-\bm{p}+\bm{q}/2,\downarrow}-i\nu_n} \notag \\
    \approx& \sum_{\bm{p}}\frac{1}{\xi_{\bm{p}+\bm{q}/2,\uparrow}+\xi_{-\bm{p}+\bm{q}/2,\downarrow}-i\nu_n}, \label{eq:PI_bec}
 \end{align}
 where we have replaced the dressed Green's function in Eq.~\eqref{eq:Greens_func} by the bare one, 
 $G_{0,\sigma}(\bm{p},i\omega_m)=[i\omega_m-\xi_{\bm{p},\sigma}]^{-1}$, for simplicity.
 In addition, since $\mu_\sigma<0$ and $|\mu_\sigma/\epsilon_\mathrm{F}|\gg 1$ in the strong-coupling regime, 
 we have dropped the Fermi distribution function $n_\mathrm{F}(x)$ in obtaining the last expression.
 Eqs.~\eqref{eq:PI_bec} and \eqref{eq:a_eff} yield the inverse of the scattering matrix in Eq.~\eqref{eq:Scattering_matrix} as
 \begin{widetext}
 \begin{align}
    \Gamma^{-1}(\bm{q},i\nu_n)=\frac{m}{4\pi a_{\mathrm{eff}}} 
    +\sum_{\bm{p}}&\left[
        \frac{1}{\xi_{\bm{p}+\bm{q}/2,\uparrow}+\xi_{-\bm{p}+\bm{q}/2,\downarrow}-i\nu_n}
    -\frac{1}{\epsilon_{\bm{p},\uparrow}+\epsilon_{-\bm{p},\downarrow}} \label{eq:Gamma_bec}
    \right].
 \end{align}
 When we expand Eq.~\eqref{eq:Gamma_bec} with respect to $1/|E_\mathrm{b}|$, where 
 \begin{align}
   E_\mathrm{b}=2(t_\uparrow+t_\downarrow)[1-\cosh[(a_\mathrm{eff}/d_z)^{-1}]] \label{eq:Eb_App}
 \end{align} 
 is the two-body binding energy, 
 the equation \eqref{eq:Gamma_bec} reduces to, up to the second order,
 \begin{align}
    \Gamma^{-1}(\bm{q},i\nu_n)=\frac{m}{4\pi d_z|E_\mathrm{b}|}\left[
      i\nu_n-\frac{\bm{q_\perp}^2}{4m}+\mu_\mathrm{B}-\frac{2t_\uparrow t_\downarrow}{|E_\mathrm{b}|}[1-\cos{(q_zd_z)}]\quad -\frac{4(t_\uparrow+t_\downarrow)^2}{|E_\mathrm{b}|} \left[
        1-\left(1+\frac{\bm{q_\perp}^2/4m-\mu_\mathrm{B}-i\nu_n}{2(t_\uparrow+t_\downarrow)}\right)^2
        \right]
        \right],
 \end{align}
 \end{widetext}
 where $\mu_{\mathrm{B}}=\mu_\uparrow+\mu_\downarrow-E_{\mathrm{b}}$.
 Retaing the leading contributions with respect to $q_\perp$ and $[1-\cos(q_zd_z)]$, we have
 \begin{align}
    \Gamma(\bm{q},i\nu_n)
        =\frac{4\pi d_z|E_\mathrm{b}|}{m}\frac{1}
            {i\nu_n-\epsilon_{\mathrm{B},\bm{q}}+\mu_\mathrm{B}}, \label{eq:Gamma_bec2}
 \end{align}
 where 
 \begin{align}
    \epsilon_{\mathrm{B},\bm{q}}&=\frac{q_\perp^2}{4m}+t_{\mathrm{B},z}[1-\cos(q_zd_z)], \label{eq:Ap_eps_B}\\
    t_{\mathrm{B},z}&=\frac{2t_\uparrow t_\downarrow d_z^2}{|E_\mathrm{b}|}.
 \end{align}

 We note that $E_\mathrm{b}$ in Eq.~\eqref{eq:Eb_App} is obtained by solving the two-body Schrödinger equation
 \begin{align}
    &\left[
        -\frac{\nabla_{\bm{r}_1}^2}{2m}-\frac{\nabla_{\bm{r}_2}^2}{2m}-U\delta(\bm{r}_1-\bm{r}_2) \right.\notag \\
        &\left.
        +V_{\rm{OL}}^{\uparrow}(r_{1z})+V_{\rm{OL}}^{\downarrow}(r_{2z})
        \right]\psi(\bm{r}_1,\bm{r}_2)
        =E\psi(\bm{r}_1,\bm{r}_2),
 \end{align}
 under the tight-binding approximation~\cite{RevModPhys.80.885}, 
 where $V_{\rm{OL}}^{\uparrow}(r_{1z})$ and $V_{\rm{OL}}^{\downarrow}(r_{2z})$ are species-selective optical lattice potentials, 
 satisfying $V_{\rm{OL}}^{\sigma}(r_{z})=V_{\rm{OL}}^{\sigma}(r_{z}+d_z)$.
 The reason we do not obtain a standard three-dimensional expression~\cite{Randeria2014,OHASHI2020103739,RevModPhys.80.1215}
 \begin{align}
    E_\mathrm{b}=-\frac{1}{ma_s^2}, \label{eq:E_b_2D}
 \end{align}
 is that the ultraviolet divergence in this mixed-dimensional system is regularized using the ``3D-like'' effective scattering length~\cite{PhysRevA.82.063628} in Eq.~\eqref{eq:a_eff}.
 If we remove this divergence by using the ``2D-like'' effective scattering length $a_\mathrm{eff}^{\mathrm{2D}}$
 \begin{align}
    \frac{1}{U}=&\frac{m}{2\pi d_z}\ln(k_\mathrm{F}a_\mathrm{eff}^{\mathrm{2D}}) 
    \notag \\&+ 
    \int_{-\pi/d_z}^{\pi/d_z}\frac{\dd{p_z}}{2\pi}
    \int_{k_\mathrm{F}}^{\infty}\frac{\dd{p_\perp}}{(2\pi)^2} 
    2\pi p_\perp\frac{1}{\epsilon_{\bm{p},\uparrow}+\epsilon_{\bm{p},\downarrow}},
 \end{align}
 we recover Eq.~\eqref{eq:E_b_2D} with $a_s$ replaced by $a_\mathrm{eff}^{\mathrm{2D}}$.

 We also note that the coefficient of $[1-\cos(q_zd_z)]$ in Eq.~\eqref{eq:Ap_eps_B}, 
 which corresponds to the effective molecular hopping parameter of the bosonic molecules, 
 differs from $t_{\mathrm{B},z}=4t_\uparrow t_\downarrow d_z^2/|E_\mathrm{b}|$, 
 which obtained in the strong-coupling BEC regime of the standard 3D attractive Hubbard model ~\cite{RevModPhys.80.885,PhysRev.115.2,RevModPhys.62.113}. 
 This discrepancy originates from the fact that, while the tight binding model is used in all the $x,y,z$ directions in the 3D Hubbard model, 
 Eq.~\eqref{eps} has the continuous free-particle dispersion in the in-plane direction.

 From Eq.~\eqref{eq:Gamma_bec2} and \eqref{eq:Number}, we obtain 
 \begin{align}
    N&=T\sum_{\bm{p},i\omega_m,\sigma}G^0_\sigma(\bm{p},i\omega_m)\Sigma^\mathrm{T}_\sigma(\bm{p},i\omega_m)G^0_\sigma(\bm{p},i\omega_m) \notag \\
    &=-T\sum_{\bm{q},i\nu_n,\sigma}\frac{\partial}{\partial\mu_\sigma}\log\left[
        \Gamma^{-1}(\bm{q},i\nu_n)
    \right] \notag \\
    &=-T\sum_{\bm{q},i\nu_n,\sigma}\frac{1}{i\nu_n-\epsilon_{\mathrm{B},\bm{q}}+\mu_\mathrm{B}}\notag\\
    &=2\sum_{\bm{q}}n_\mathrm{B}(\epsilon_{\mathrm{B},\bm{q}}-\mu_\mathrm{B}), \label{eq:number_bose_App}
 \end{align}
 where we evaluated Eq.~\eqref{eq:Number} by expanding the Dyson equation and 
 retaining only the first-order term with respect to the non-self-consistent $T$-matrix self-energy,
 \begin{align}
   \Sigma^\mathrm{T}_\sigma(\bm{p},i\omega_m)=T\sum_{\bm{q},i\nu_n}&
   \Gamma(\bm{q},i\nu_n)
   \notag\\\times&
   G_{-\sigma}^0(\bm{q-p},i\nu_n-i\omega_m)e^{\delta\cdot i\nu_n},
 \end{align}
 which provides the dominant contribution in the strong-coupling regime~\cite{OHASHI2020103739}.
 When the BEC transition occurs, $\mu_\mathrm{B}$ approaches zero, and 
 Eq.~\eqref{eq:number_bose_App} reduces to Eq.~\eqref{eq:number_bose} with $N_\mathrm{B}=N/2$.

 Carrying out the $q_\perp$-integration in Eq.~\eqref{eq:number_bose_App}, one has
 \begin{align}
    N_\mathrm{B}&=-\frac{md_z}{\pi\beta}\int_{-\pi/d_z}^{\pi/d_z}\frac{\dd{q_z}}{2\pi}\ln[1-e^{-\beta t_{\mathrm{B},z}[1-\cos(q_zd_z)]}] \notag \\
    &=\frac{md_z}{\pi\beta}\sum_{n=1}^{\infty}\frac{e^{-\beta t_{\mathrm{B},z}n}}{n}
    \int_{-\pi/d_z}^{\pi/d_z}\frac{\dd{q_z}}{2\pi} e^{\beta t_{\mathrm{B},z}n\cos(q_zd_z)}\notag\\
    &=\frac{md_z}{\pi\beta}\sum_{n=1}^{\infty}\frac{e^{-\beta t_{\mathrm{B},z}n}}{n}I_0(\beta t_{\mathrm{B},z}n), 
 \end{align}
 where $I_0(x)$ is the modified Bessel function of the first kind.
 Assuming that $\beta t_{\mathrm{B},z} \ll 1$, one can approximate $I_0(\beta t_{\mathrm{B},z}n)$ to unity. 
 We then reach
 \begin{align}
    T_{\mathrm{c}}=\frac{\pi N_\mathrm{B}}{md_z}\left[
        W\left(\frac{\pi N_\mathrm{B}}{md_z t_{\mathrm{B},z}}\right)
    \right]^{-1}, \label{eq:Ap_T_BEC}
 \end{align}
 where $W(Y)$ which satisfies $Y=Xe^X, X=W(Y)$ is the Lambert W function.
 The first line in Eq.~\eqref{eq:T_BEC} is obtained by substituting $t_{\mathrm{B},z}=2t_\uparrow t_\downarrow/|E_\mathrm{b}|$ into Eq.~\eqref{eq:Ap_T_BEC}.
 We briefly note that Eq.~\eqref{eq:Ap_T_BEC} is consistent with the assumption $\beta t_{\mathrm{B},z} \ll 1$.
 \begin{figure}[t]
   \centering
   \includegraphics[width=0.9\linewidth]{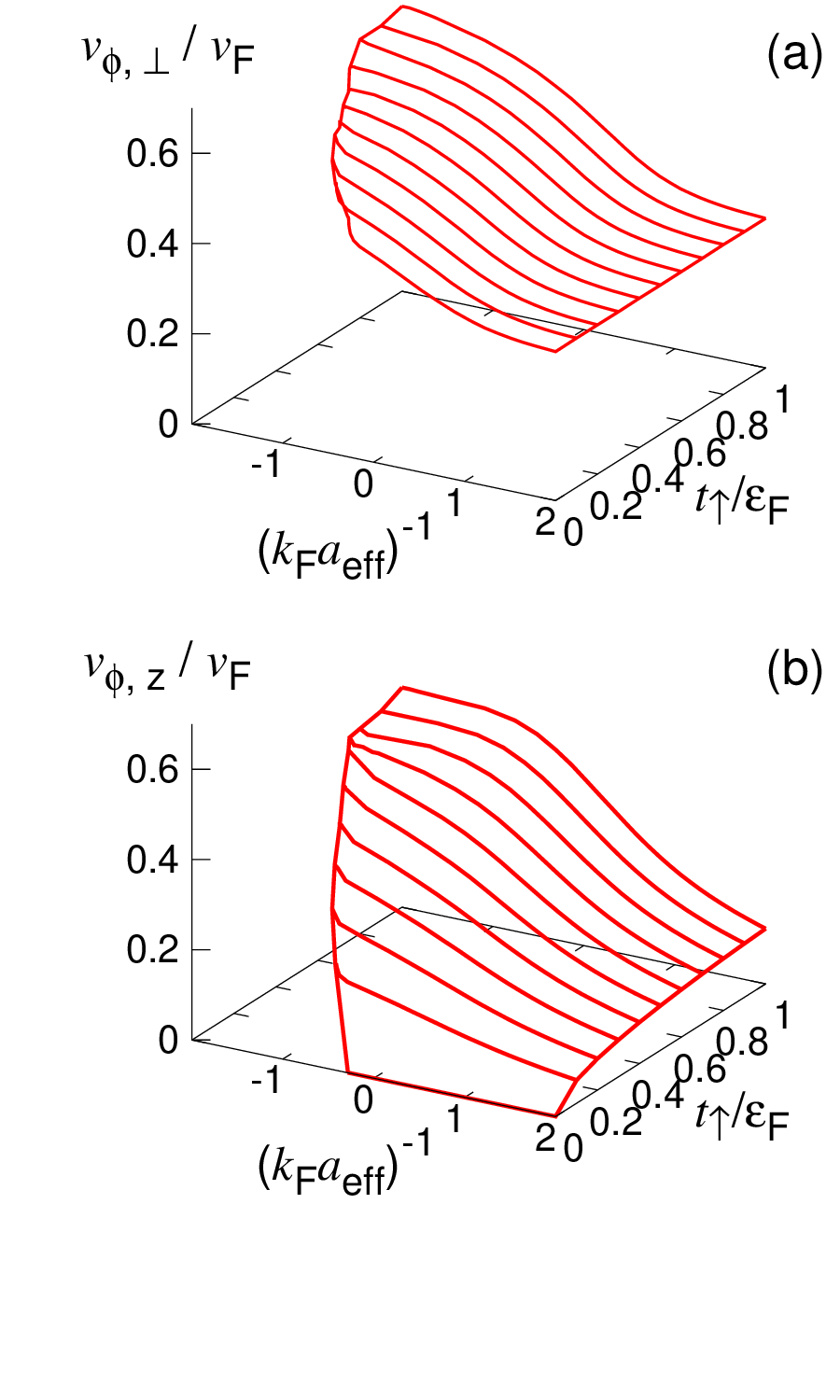}
   \vspace*{-2cm}
   \caption{
       Calculated velocity $\bm{v}_{\phi}=(\bm{v}_{\phi,\perp},v_{\phi,z})$ of the Goldstone mode at $T=0$, within the framework of GRPA.
       (a) In-plane component $v_{\phi,\perp}$.
       (b) Perpendicular component $v_{\phi,z}$.
       $v_\mathrm{F}=k_\mathrm{F}/m$.
   }
   \label{fig:v_ng_perp_z}
 \end{figure}

 \section{GRPA calculation for the velocity of the Goldstone mode} \label{sec:GRPA}
 To evaluate the velocity $\bm{v}_{\phi}=(\bm{v}_{\phi,\perp},v_{\phi,z})$ of the Goldstone mode shown in Fig.~\ref{fig:v_ng_perp_z}, 
 we consider the following generalized correlation functions in the Cooper channel~\cite{doi:10.1143/JPSJ.66.2437}: 
 \begin{align}
    \Pi_{ij}(\bm{q},\omega)=-i\int_0^\infty \dd{t} e^{-i\omega t}\Braket{[\rho_{i,\bm{q}}(t),\rho_{j,-\bm{q}}(0)]}, \label{eq:corelation_heisenberg}
 \end{align}
 where $i,j=1,2$ and 
 $\rho_{i,\bm{q}}(t)$ is the Heisenberg representation of the generalized density operator in Eq.~\eqref{eq:rho_123}.
 Since the Goldstone mode in the superfluid phase is a collective phase oscillation of the superfluid order parameter, 
 it appears as a pole of the phase-phase correlation function $\Pi_{22}(\bm{q},\omega)$.
 Thus, from the $\bm{q}$-dependence of the pole $\omega(\bm{q})$ around $\bm{q}=0$, 
 the mode velocity $\bm{v}_{\phi}$ is obtained through the relation, 
 $\omega(\bm{q})=\bm{v}_\phi\cdot\bm{q}$.

 The pair-correlation function in Eq.~\eqref{eq:corelation_heisenberg} can conveniently be evaluated by way of the corresponding correlation function in the Matsubara formalism, given by
 \begin{align}
    \Pi_{ij}(\bm{q},i\nu_n)=-\int_{0}^{\beta}\dd{\tau} e^{-i\nu_n\tau}\Braket{T_\tau\{\rho_{i,\bm{q}}(\tau)\rho_{j,-\bm{q}}(0)\}},
 \end{align}
 which is related to $\Pi_{ij}(\bm{q},\omega)$ in Eq.~\eqref{eq:corelation_heisenberg} as
 \begin{align}
    \Pi_{ij}(\bm{q},\omega+i\delta)=\Pi_{ij}(\bm{q},i\nu_n\rightarrow \omega+i\delta).
 \end{align}
 Evaluating $\Pi_{ij}(\bm{q},i\nu_n)$ within the generalized random phase approximation (GRPA), 
 we obtain~\cite{doi:10.1143/JPSJ.66.2437},  
 \begin{align}
    \Pi_{22}(\bm{q},i\nu_n)&=\frac{\tilde{\Pi}_{22}(\bm{q},i\nu_n)}{1+\frac{U}{2}\tilde{\Pi}_{22}(\bm{q},i\nu_n)}. \label{eq:A12}
 \end{align}
 Here
 \begin{align}
    \tilde{\Pi}_{22}(\bm{q},i\nu_n)=&\Pi^0_{22}(\bm{q},i\nu_n)
    \notag \\&
    +\Pi^0_{21}(\bm{q},i\nu_n)
    \frac{-\frac{U}{2}}{1+\frac{U}{2}\Pi^0_{11}(\bm{q},i\nu_n)}\Pi^0_{12}(\bm{q},i\nu_n),\label{eq:A13}
 \end{align}
 where
 \begin{align}
    \Pi^0_{ij}(\bm{q},i\nu_n)=T\sum_{\bm{p},i\omega_m}\Tr&\left[
        \tau_i \hat{G}^0(\bm{p}+\bm{q}/2,i\omega_m+i\nu_n)
        \right.\notag\\&\times\left.
        \tau_j \hat{G}^0(\bm{p}-\bm{q}/2,i\omega_m)
    \right], \label{eq:PI0_ij}
 \end{align}
 is the mean-field pair-correlation function (when $\hat{G}^0(\bm{p},i\omega_m)$ is given in Eq.~\eqref{eq:G0}).
 Carrying out the Matsubara frequency summation in Eq.~\eqref{eq:PI0_ij}, 
 one has 
 \begin{widetext}
 \begin{align}
    \Pi^0_{11}(\bm{q},i\omega_m)&= -\sum_{\bm{p}}\left[
        1+\frac{\xi_{+,\bm{p}+\frac{\bm{q}}{2}}\xi_{+,\bm{p}-\frac{\bm{q}}{2}}-\Delta^2}{E_{\bm{p}+\frac{\bm{q}}{2}}E_{\bm{p}-\frac{\bm{q}}{2}}}
    \right]
    \frac{E_{\bm{p}+\frac{\bm{q}}{2}}+E_{\bm{p}-\frac{\bm{q}}{2}}+\xi_{-,\bm{p}+\frac{\bm{q}}{2}}-\xi_{-,\bm{p}-\frac{\bm{q}}{2}}}{(E_{\bm{p}+\frac{\bm{q}}{2}}+E_{\bm{p}-\frac{\bm{q}}{2}}+\xi_{-,\bm{p}+\frac{\bm{q}}{2}}-\xi_{-,\bm{p}-\frac{\bm{q}}{2}})^2+\omega_m^2},\label{eq:A15}\\
    \Pi^0_{22}(\bm{q},i\omega_m)&= -\sum_{\bm{p}}\left[
        1+\frac{\xi_{+,\bm{p}+\frac{\bm{q}}{2}}\xi_{+,\bm{p}-\frac{\bm{q}}{2}}+\Delta^2}{E_{\bm{p}+\frac{\bm{q}}{2}}E_{\bm{p}-\frac{\bm{q}}{2}}}
    \right]
    \frac{E_{\bm{p}+\frac{\bm{q}}{2}}+E_{\bm{p}-\frac{\bm{q}}{2}}+\xi_{-,\bm{p}+\frac{\bm{q}}{2}}-\xi_{-,\bm{p}-\frac{\bm{q}}{2}}}{(E_{\bm{p}+\frac{\bm{q}}{2}}+E_{\bm{p}-\frac{\bm{q}}{2}}+\xi_{-,\bm{p}+\frac{\bm{q}}{2}}-\xi_{-,\bm{p}-\frac{\bm{q}}{2}})^2+\omega_m^2},\label{eq:A16}\\
    \Pi^0_{12}(\bm{q},i\omega_m)&= -\sum_{\bm{p}}\left[
        \frac{\xi_{+,\bm{p}-\frac{\bm{q}}{2}}}{E_{\bm{p}-\frac{\bm{q}}{2}}}
        +
        \frac{\xi_{+,\bm{p}+\frac{\bm{q}}{2}}}{E_{\bm{p}+\frac{\bm{q}}{2}}}
    \right]
    \frac{\omega_m}{(E_{\bm{p}+\frac{\bm{q}}{2}}+E_{\bm{p}-\frac{\bm{q}}{2}}+\xi_{-,\bm{p}+\frac{\bm{q}}{2}}-\xi_{-,\bm{p}-\frac{\bm{q}}{2}})^2+\omega_m^2},\label{eq:A17}\\
    \Pi^0_{21}(\bm{q},i\omega_m)&= -\Pi^0_{12}(\bm{q},i\omega_m)\label{eq:A18}.
 \end{align}
 \end{widetext}

 Substituting Eqs.~\eqref{eq:A15}-\eqref{eq:A18} into Eq.~\eqref{eq:A12} and \eqref{eq:A13}, 
 which is followed by the analytic continuation, $i\omega_m\rightarrow\omega+i\delta$, 
 we reach the GRPA expression for the phase-phase correlation function $\Pi_{22}(\bm{q},\omega)$.
 Looking for the pole energy $\omega(\bm{q})=\bm{v}_\phi\cdot\bm{q}=v_{\phi,\perp}q_\perp+v_{\phi,z}q_z$ in the low momentum limit, 
 we obtain 
 \begin{align}
    v_{\phi,\perp}&=\sqrt{\frac{\sum_{\bm{p}}\left[\frac{3p^2\Delta^2}{4E_{\bm{p}}^5}+\frac{\xi_{+,\bm{p}}}{4E_{\bm{p}}^3}\right]}{\sum_{\bm{p}}\frac{1}{E_{\bm{p}}^3}+\frac{\left(\sum_{\bm{p}}\xi_{+,\bm{p}}E_{\bm{p}}^{-3}\right)^2}{\sum_{\bm{p}}(\Delta^2 / E_{\bm{p}}^{3})}}}v_\mathrm{F}, \label{eq:v_ng_perp}\\
    v_{\phi,z}&=\sqrt{\frac{\sum_{\bm{p}}\left[\frac{3\Delta^2A^2}{2E_{\bm{p}}^5}+\frac{\xi_{+,\bm{p}}B}{E_{\bm{p}}^3}-\frac{C}{4E_{\bm{p}}^3}\right]}{\sum_{\bm{p}}\frac{1}{E_{\bm{p}}^3}+\frac{\left(\sum_{\bm{p}}\xi_{+,\bm{p}}E_{\bm{p}}^{-3}\right)^2}{\sum_{\bm{p}}(\Delta^2 / E_{\bm{p}}^3)}}}v_\mathrm{F}, \label{eq:v_ng_z}
 \end{align}
 where 
 \begin{align}
    A&=\frac{t_\uparrow+t_\downarrow}{2}\sin (p_zd_z),\\
    B&=\frac{t_\uparrow+t_\downarrow}{8}\cos (p_zd_z),\\
    C&=(t_\uparrow-t_\downarrow)\sin(p_zd_z),
 \end{align}
 and $v_\mathrm{F}=k_\mathrm{F}/m$.
 GRPA is consistent with the mean-field BCS theory in the sense that substitution of the latter solution into the former guarantees the required gapless dispersion of the Goldstone mode.
 Keeping this in mind, using the self-consistent solutions ($\Delta, \mu_\uparrow, \mu_\downarrow$) obtained in Appendix \ref{sec:mf_BCS} to numerically evaluate Eqs.~\eqref{eq:v_ng_perp} and \eqref{eq:v_ng_z}, 
 we obtain Figs.~\ref{fig:v_ng_2D3D} and \ref{fig:v_ng_perp_z}. 

 We find from Fig.~\ref{fig:v_ng_perp_z}(b) that $v_{\phi,z}$ always vanishes in the 2D-3D limit: 
 When $t_\uparrow=0$, the numerator inside the square root of Eq.~\eqref{eq:v_ng_z} is reduced to 
 \begin{align}
    &t_\downarrow^2\sum_{\bm{p}}\left[
      \left(\frac{3\Delta^2}{8E_{\bm{p}}^5}-\frac{1}{4E_{\bm{p}}^3}\right)\sin^2 (p_zd_z)
      +
      \frac{\xi_{+,\bm{p}}}{8E_{\bm{p}}^3}\cos^2 (p_zd_z)
  \right]\notag \\
    =&\frac{t_\downarrow^2}{8}\sum_{\bm{p}_\perp}
    \int_{-\pi/d_z}^{\pi/d_z} \frac{\dd{p_z}}{2\pi}
    \frac{\partial}{\partial p_z}\frac{\xi_{+,\bm{p}}\sin (p_zd_z)}{E_{\bm{p}}^3} \notag\\
    =&\frac{t_\downarrow^2}{16\pi}\sum_{\bm{p}_\perp}\left.
        \frac{\xi_{+,\bm{p}}\sin (p_zd_z)}{E_{\bm{p}}^3}\right|_{p_z=-\pi/d_z}^{p_z=\pi/d_z}\notag\\
    =&0. \label{eq:B1}
 \end{align}
 On the other hand, the denominator inside the square root of Eq.~\eqref{eq:v_ng_z} remains non-zero, 
 $v_{\phi,z}=0$ is obtained, 
 irrespective of the interaction strength.

 \bibliographystyle{apsrev4-2}

\end{document}